\documentclass[11pt,a4paper]{article}
\usepackage{jheppub}

\usepackage{graphicx}
\usepackage{tensor}
\usepackage{multirow}
\usepackage{xcolor}
\usepackage{amsmath}
\usepackage{mathtools}

\graphicspath{{}{images/}{tikz/}}

\newcommand{\elp}{e^+}
\newcommand{\elm}{e^-}
\newcommand{\pip}{\pi^+}
\newcommand{\pim}{\pi^-}
\newcommand{\piz}{\pi^0}
\newcommand{\kp}{K^+}
\newcommand{\km}{K^-}
\newcommand{\ks}{K^0_S}
\newcommand{\pr}{p}
\newcommand{\apr}{\bar{p}}
\newcommand{\lam}{\Lambda}
\newcommand{\alam}{\bar{\Lambda}}

\newcommand{\Dz}{D^0}

\newcommand{\Bp}{B^+}
\newcommand{\Bm}{B^-}
\newcommand{\Bz}{B^0}

\newcommand{\decaykp}{B^+ \to \eta_{c2}(1D) \kp}
\newcommand{\decayks}{B^0 \to \eta_{c2}(1D) \ks}
\newcommand{\decaypimk}{B \to \eta_{c2}(1D) \pi K}
\newcommand{\decaypimkp}{B^0 \to \eta_{c2}(1D) \pim \kp}
\newcommand{\decaypimks}{B^+ \to \eta_{c2}(1D) \pip \ks}

\newcommand{\invfb}{\text{fb}^{-1}}
\newcommand{\cm}{\text{cm}}
\newcommand{\kev}{\text{keV}}
\newcommand{\mev}{\text{MeV}}
\newcommand{\mevcc}{\text{MeV}/c^2}
\newcommand{\gev}{\text{GeV}}

\newcommand{\gevcc}{\text{GeV}/c^2}

\newcommand{\mbc}{M_{\text{bc}}}
\newcommand{\de}{\Delta E}

\newcommand{\etaccha}[1]{
\ifcase#1\relax
\kp \ks \pim
\or
\kp \km \piz
\or
\ks \ks \piz
\or
\kp \km \eta
\or
\kp \km \eta_{3 \pi}
\or
\kp \km \kp \km
\or
\kp \km \pip \pim
\or
\kp \km \pip \pim \piz
\or
\ks \km \pip \pim \pip
\or
\eta' (\to \eta \pip \pim) \pip \pim
\or
\eta' (\to \eta_{3 \pi} \pip \pim) \pip \pim 
\or
\eta_{2 \gamma} \pip \pim
\or
\eta_{3 \pi} \pip \pim
\or
\pr \apr
\or
\pr \apr \piz
\or
\pr \apr \pip \pim
\or
\lam \alam
\or
\hccha{1}
\fi
}
\newcommand{\etacdec}[1]{\eta_c \to \etaccha{#1}}
\newcommand{\hccha}[1]{
\ifcase#1\relax
\eta_c \gamma
\or
\pr \apr \pip \pim
\fi
}
\newcommand{\hcdec}[1]{h_c \to \hccha{#1}}
\newcounter{optchannel}
\newcommand{\optcha}[1]{
\ifnum#1<17
\etaccha{#1}
\else
{
\setcounter{optchannel}{#1}
\addtocounter{optchannel}{-16}
\hccha{\arabic{optchannel}}
}
\fi
}

\newcommand{\br}{\mathcal{B}}

\begin{document}

\title{First search for the $\eta_{c2}(1D)$ in $B$ decays at Belle}
\collaborationImg{\includegraphics[width=2cm]{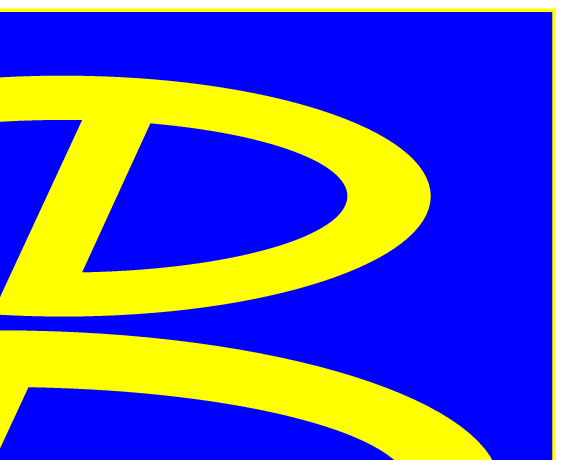}}
\collaboration{The Belle collaboration}
\author[43]{K.~Chilikin}
\author[18,14]{I.~Adachi}
\author[85]{H.~Aihara}
\author[78,37]{S.~Al~Said}
\author[4]{D.~M.~Asner}
\author[5,63]{V.~Aulchenko}
\author[34]{T.~Aushev}
\author[78]{R.~Ayad}
\author[9]{V.~Babu}
\author[23]{S.~Bahinipati}
% ...
\affiliation[4]{Brookhaven National Laboratory, Upton, New York 11973, USA}
\affiliation[5]{Budker Institute of Nuclear Physics SB RAS, Novosibirsk 630090, Russia}
% ...
\affiliation[9]{Deutsches Elektronen--Synchrotron, 22607 Hamburg, Germany}
% ...
\affiliation[14]{SOKENDAI (The Graduate University for Advanced Studies), Hayama 240-0193, Japan}
% ...
\affiliation[18]{High Energy Accelerator Research Organization (KEK), Tsukuba 305-0801, Japan}
% ...
\affiliation[23]{Indian Institute of Technology Bhubaneswar, Satya Nagar 751007, India}
% ...
\affiliation[34]{Higher School of Economics (HSE), Moscow 101000, Russia}
% ...
\affiliation[37]{Department of Physics, Faculty of Science, King Abdulaziz University, Jeddah 21589, Saudi Arabia}
% ...
\affiliation[43]{P.N. Lebedev Physical Institute of the Russian Academy of Sciences, Moscow 119991, Russia}
% ...
\affiliation[63]{Novosibirsk State University, Novosibirsk 630090, Russia}
% ...
\affiliation[78]{Department of Physics, Faculty of Science, University of Tabuk, Tabuk 71451, Saudi Arabia}
% ...
\affiliation[85]{Department of Physics, University of Tokyo, Tokyo 113-0033, Japan}
% ...

\abstract{
The first dedicated search for the $\eta_{c2}(1D)$ is carried out
using the decays $\decaykp$, $\decayks$, $\decaypimkp$, and $\decaypimks$
with $\eta_{c2}(1D) \to h_c \gamma$.
No significant signal is found. For the $\eta_{c2}(1D)$ mass range
between $3795$ and $3845\ \mevcc$, the branching-fraction upper limits are
determined to be
$\br(\decaykp) \times \br(\eta_{c2}(1D) \to h_c \gamma) < 3.7 \times 10^{-5}$,
$\br(\decayks) \times \br(\eta_{c2}(1D) \to h_c \gamma) < 3.5 \times 10^{-5}$,
$\br(\decaypimkp) \times \br(\eta_{c2}(1D) \to h_c \gamma)
 < 1.0 \times 10^{-4}$, and
$\br(\decaypimks) \times \br(\eta_{c2}(1D) \to h_c \gamma)
 < 1.1 \times 10^{-4}$ at 90\% C. L.
The analysis is based on the 711 $\invfb$ data sample collected
on the $\Upsilon(4S)$ resonance by the Belle detector,
which operated at the KEKB asymmetric-energy $\elp \elm$ collider.
}



\maketitle

\section{Introduction}

The first observed $D$-wave charmonium state was the $\psi(3770)$,
which can be produced in $\elp \elm$ collisions directly~\cite{Rapidis:1977cv}
(note that the $\psi(3770)$ is predominantly $1^3D_1$ state,
but includes an admixture of $S$-wave vector charmonia~\cite{Godfrey:1985xj}).
Other $D$-wave charmonium states can be produced in $b$-hadron decays,
hadronic transitions from other charmonium states,
or directly in hadron collisions.
Evidence for the $\psi_2(1D)$ was first found by the Belle
collaboration in the decays $\Bp \to \psi_2(1D)
(\to \chi_{c1} (\to J/\psi \gamma) \gamma) \kp$~\cite{Bhardwaj:2013rmw};
this state was later observed by BESIII in
the process $e^+ e^- \to \psi_2(1D) (\to \chi_{c1} \gamma)
\pip \pim$~\cite{Ablikim:2015dlj}.
A candidate for the $\psi_3(1D)$ state, the $X(3842)$,
was recently observed by the LHCb collaboration using direct production
in $p p$ collisions~\cite{Aaij:2019evc}.

The $1^1D_2$ charmonium state, the $\eta_{c2}(1D)$, has not been observed
experimentally yet.
Various potential models~\cite{Godfrey:1985xj, Fulcher:1991dm, Zeng:1994vj,
Ebert:2002pp} predict that the masses of the $\eta_{c2}(1D)$
and $\psi_2(1D)$ are very close to each other
(see also the summary table in ref.~\cite{Brambilla:2004wf}).
While the predicted $\eta_{c2}(1D)$ mass can vary by up to
$70\ \mevcc$ between models,
the typical value of the mass difference between
the $\eta_{c2}(1D)$ and $\psi_2(1D)$ in a given model is about 2 $\mevcc$.
Lattice calculations~\cite{Mohler:2012na, Cheung:2016bym} also find
that the $\eta_{c2}(1D)$ and $\psi_2(1D)$ masses are close to each other.
The mass difference calculated from the results of ref.~\cite{Cheung:2016bym}
is $m_{\eta_{c2}(1D)} - m_{\psi_2(1D)} = 9 \pm 10\ \mevcc$
assuming uncorrelated uncertainties, where the uncertainty is statistical only.
In addition, the hyperfine splitting of the 1D charmonium states
is expected to be small; one can use the known masses of the $1^3D_J$ states
to estimate the expected $\eta_{c2}(1D)$ mass: 
$m_{\eta_{c2}(1D)} \approx (3 m_{\psi(3770)} + 5 m_{\psi_2(1D)} +
7 m_{X(3842)}) / 15 \approx 3822\ \mevcc$.
Thus, the $\eta_{c2}(1D)$ mass is expected to be around $3820\ \mevcc$,
which is below the $D^*\bar{D}$ threshold.
The decay $\eta_{c2}(1D) \to D \bar{D}$ is forbidden by parity conservation.
The $\eta_{c2}(1D)$ remains the only unobserved
conventional charmonium state that does not have open-charm decays.

The $\eta_{c2}(1D)$ is expected to decay predominantly
via an E1 transition to $h_c \gamma$.
The partial decay widths were estimated in ref.~\cite{Eichten:2002qv} to be
$\Gamma(\eta_{c2}(1D) \to h_c \gamma) = 303\ \kev$,
$\Gamma(\eta_{c2}(1D) \to g g) = 110\ \kev$,
and $\Gamma(\eta_{c2}(1D) \to \eta_c \pi \pi) = 45\ \kev$, resulting in
a $\br(\eta_{c2}(1D) \to h_c \gamma)$ branching fraction of about 0.7.
Other estimates of $\Gamma(\eta_{c2}(1D) \to h_c \gamma)$ are in the range
of $\sim(250 - 350)\ \kev$~\cite{Barnes:2005pb,Ebert:2002pp}.
Another calculation of the width of the $\eta_{c2}(1D)$ decays to light hadrons
was carried out in ref.~\cite{Fan:2009cj}; the resulting estimate of the
branching fraction $\br(\eta_{c2}(1D) \to h_c \gamma) = (44 - 54)\%$ is
somewhat lower, but the radiative decay is still expected to be dominant.
The branching fraction of the decay $\decaykp$ has been calculated using
the rescattering mechanism to be
$(1.72 \pm 0.42) \times 10^{-5}$~\cite{Xu:2016kbn}.
If this prediction is correct, then the branching-fraction product
$\br(\decaykp) \times \br(\eta_{c2}(1D) \to h_c \gamma)$ is expected to be
about $1.0 \times 10^{-5}$.

Recently the Belle collaboration found evidence for the decay
$\Bp \to h_c \kp$~\cite{Chilikin:2019wzy}. Its final state is very similar
to that of the $\decaykp$ decay. Thus, the analysis method used in
ref.~\cite{Chilikin:2019wzy} can be applied to the $\eta_{c2}(1D)$ search
with minor modifications.
Here we present a search for the $\eta_{c2}(1D)$ using the
decays $\decaykp$, $\decayks$, $\decaypimkp$, and $\decaypimks$
with $\eta_{c2}(1D) \to h_c \gamma$.
The analysis is performed using the $711\ \invfb$ data sample collected
by the Belle detector at the KEKB asymmetric-energy $\elp \elm$
collider~\cite{Kurokawa:2001nw, Abe:2013kxa} at the $\Upsilon(4S)$ resonance,
which contains $772 \times 10^6$ $B\bar{B}$ pairs.

\section{The Belle detector}

The Belle detector is a large-solid-angle magnetic
spectrometer that consists of a silicon vertex detector (SVD),
a 50-layer central drift chamber (CDC), an array of
aerogel threshold Cherenkov counters (ACC),
a barrel-like arrangement of time-of-flight
scintillation counters (TOF), and an electromagnetic calorimeter (ECL)
comprised of CsI(Tl) crystals located inside
a superconducting solenoid coil that provides a 1.5~T
magnetic field.  An iron flux-return located outside of
the coil is instrumented to detect $K_L^0$ mesons and to identify
muons.  The detector
is described in detail elsewhere~\cite{Abashian:2000cg, Brodzicka:2012jm}.
Two inner detector configurations were used. A 2.0 cm radius beampipe
and a 3-layer silicon vertex detector were used for the first sample
of 140 $\invfb$, while a 1.5 cm radius beampipe, a 4-layer
silicon detector and a small-cell inner drift chamber were used to record
the remaining data~\cite{Natkaniec:2006rv}.

We use a {\sc geant}-based Monte Carlo (MC) simulation~\cite{Brun:1987ma}
to model the response of the detector, identify potential backgrounds and
determine the acceptance. The MC simulation includes run-dependent
detector performance variations and background conditions.  Signal MC
events are generated with {\sc EvtGen}~\cite{evtgen}
in proportion to the relative luminosities of the
different running periods.

\section{Event selection}

We select the decays
$\decaykp$, $\decayks$, $\decaypimkp$, and $\decaypimks$
with $\eta_{c2}(1D) \to h_c \gamma$ and $h_c \to \eta_c \gamma$.
The $\eta_c$ candidates are reconstructed in ten different decay channels
as described below.
Inclusion of charge-conjugate modes is implied hereinafter.
The reconstruction uses the Belle II analysis framework~\cite{Kuhr:2018lps}
with a conversion from Belle to Belle II data format~\cite{Gelb:2018agf}.

All tracks are required to originate from the interaction-point region:
we require $dr < 0.2\ \cm$ and $|dz| < 2\ \cm$, where
$dr$ and $dz$ are the cylindrical coordinates of the point of the
closest approach of the track to the beam axis
(the $z$ axis of the laboratory reference frame coincides with
the positron-beam axis).

Charged $\pi$, $K$ mesons and protons are identified using
likelihood ratios
$R_{h_1/h_2} = \mathcal{L}_{h_1}/(\mathcal{L}_{h_1}+\mathcal{L}_{h_2})$,
where $h_1$ and $h_2$ are the particle-identification hypotheses ($\pi$, $K$,
or $p$) and $\mathcal{L}_{h_i}$ ($i = 1, 2$) are
their corresponding likelihoods.
The likelihoods are calculated by combining
time-of-flight information from the TOF, the number of photoelectrons from
the ACC and $dE/dx$ measurements in the CDC.
We require $R_{K/\pi} > 0.6$ and $R_{p/K} < 0.9$ for $K$ candidates,
$R_{\pi/K} > 0.1$ and $R_{p/\pi} < 0.9$ for $\pi$ candidates,
and $R_{p/\pi} > 0.6$, $R_{p/K} > 0.6$ for $p$ candidates.
The identification efficiency of the above requirements varies in the ranges
(95.0 -- 99.7)\%, (86.9 -- 94.6)\%, and (90.2 -- 98.3)\%
for $\pi$, $K$, and $p$, respectively,
depending on the $\eta_c$ decay channel.
The misidentification probability
for background particles that are not a $\pi$, $K$, and $p$,
varies in the ranges
(30 -- 48)\%, (2.1 -- 4.5)\%, and (0.6 -- 2.0)\%,
respectively.

Electron candidates are identified as CDC charged tracks
that are matched to electromagnetic showers in the ECL.
The track and ECL cluster matching quality,
the ratio of the electromagnetic shower energy to the track momentum,
the transverse shape of the shower, the ACC light yield, and the
track $dE/dx$ ionization are used in our electron identification
criteria. A similar likelihood ratio is constructed:
$R_e = \mathcal{L}_e / (\mathcal{L}_e + \mathcal{L}_h)$,
where $\mathcal{L}_e$ and $\mathcal{L}_h$ are the likelihoods for electrons and
charged hadrons ($\pi$, $K$ and $p$), respectively~\cite{Hanagaki:2001fz}.
An electron veto ($R_e < 0.9$) is imposed on $\pi$, $K$, and $p$ candidates.
This veto is not applied to the $\ks$ and $\Lambda$ daughter tracks, because
they have independent selection criteria. For $\eta_c$ decay
channels other than $\etacdec{2}$ and $\etacdec{16}$, the electron veto
rejects (2.3-17)\% of the background events, while its signal
efficiency is 97.5\% or greater.

Photons are identified as ECL electromagnetic showers that
have no associated charged tracks detected in the CDC. The
shower shape is required to be consistent with that of a photon.
The photon energies in the laboratory frame are
required to be greater than $30\ \mev$.
The photon energies in MC simulation are corrected to take
into account the difference of resolution in data and MC. Correction
factors are based on analysis of mass resolutions in the channels
$\piz \to \gamma \gamma$, $\eta \to \gamma \gamma$, and
$D^{*0} \to D^0 \gamma$~\cite{Tamponi:2015xzb}.

The $\piz$ and $\eta$ candidates are reconstructed via their
decays to two photons. The $\piz$ invariant mass is required to satisfy
$|M_{\piz} - m_{\piz}| < 15\ \mevcc$; the $\eta$ mass 
is $|M_{\eta} - m_{\eta}| < 30\ \mevcc$. Here and elsewhere,
$M_\text{particle}$ denotes the reconstructed invariant mass of the specified
particle and $m_\text{particle}$ stands for its nominal
mass~\cite{Tanabashi:2018oca}.
The requirements correspond approximately
to $3\sigma$ and $2.5\sigma$ mass windows around the nominal mass for the
$\piz$ and $\eta$, respectively. The $\eta$ decay to $\pip\pim\piz$
was also used in ref.~\cite{Chilikin:2019wzy}.
Initially, this channel was also reconstructed here,
but it was found that including it does not improve the expected sensitivity.

Candidate $V^0$ particles ($\ks$ and $\Lambda$)
are reconstructed from pairs of oppositely charged tracks that are
assumed to be $\pip \pim$ and $p \pim$ for $\ks$ and $\Lambda$, respectively.
We require $|M_{\ks} - m_{\ks}| < 20\ \mevcc$ and
$|M_{\Lambda} - m_{\Lambda}| < 10\ \mevcc$, corresponding
approximately to $5.5\sigma$ mass windows in both cases.
The $V^0$ candidates are selected by a neural network
using the following input
variables: the $V^0$ candidate momentum, the decay angle
(the angle between the momentum of a daughter track in the
$V^0$ rest frame and the direction of the boost from the laboratory frame to
the $V^0$ rest frame),
the flight distance in the $xy$ plane, the angle between the $V^0$
momentum and the direction from the interaction point to the $V^0$ vertex,
the shortest $z$ distance between the two daughter tracks,
their radial impact parameters, and numbers of hits in the SVD and CDC.
Another neural network is used to separate $\ks$ and $\Lambda$ candidates.
The input variables for this network are the
momenta and polar angles of the daughter tracks in the laboratory frame,
their likelihood ratios $R_{\pi/p}$, and the $V^0$ candidate invariant mass for
the $\Lambda$ hypothesis.
The $V^0$ identification efficiency varies in the ranges (82.2 -- 91.9)\%
and (85.1 -- 86.0)\% for $\ks$ and $\Lambda$, respectively,
depending on the $\eta_c$ and $B$ decay channels. The misidentification
probability for fake $V^0$ candidates is (0.5 -- 0.8)\% and (1.7 -- 2.4)\% for
$\ks$'s and $\Lambda$'s, respectively.

The $\eta'$ candidates are reconstructed in the $\eta \pip \pim$ decay mode.
The invariant mass is chosen in the range
$|M_{\eta'} - m_{\eta'}| < 15\ \mevcc$, which
corresponds to a $4\sigma$ mass window.

The $\eta_c$ candidates are reconstructed in ten decay channels:
$\etaccha{0}$, $\etaccha{1}$, $\etaccha{2}$, $\etaccha{3}$,
$\etaccha{5}$, $\etaccha{9}$,
$\etaccha{13}$, $\etaccha{14}$, $\etaccha{15}$, and $\etaccha{16}$.
The selected $\eta_c$ candidates
are required to satisfy $|M_{\eta_c} - m_{\eta_c}| < 50\ \mevcc$;
this mass-window width is about $1.6$ times the intrinsic width of the
$\eta_c$~\cite{Tanabashi:2018oca}.

The $h_c$ candidates are reconstructed in the channel $\hcdec{0}$;
the invariant-mass requirement is
$|M_{h_c} - m_{h_c}| < 50\ \mevcc$.
The channel $\hcdec{1}$ was used in ref.~\cite{Chilikin:2019wzy}, but
it cannot be used here because the mass resolution is not sufficient
to fully distinguish the $\chi_{c1}$ and $h_c$ peaks in the
$\hccha{1}$ mass spectrum; thus, this channel may contain a peaking background
from the process $B \to \psi_2(1D) (\to \chi_{c1} (\to \hccha{1}) \gamma) K$.
The $\eta_{c2}(1D)$ candidates are reconstructed in the channel
$\eta_{c2}(1D) \to h_c \gamma$;
their invariant mass is not restricted.

The $B$-meson candidates are reconstructed via the decay modes
$\decaykp$, $\decayks$, $\decaypimkp$, and $\decaypimks$.
The $B$ candidates are selected by
their energy and the beam-energy-constrained mass. The difference of the
$B$-meson and beam energies is defined as $\de = \sum_i E_i - E_{\text{beam}}$,
where $E_i$ are the energies of the $B$ decay products in the center-of-mass
frame and $E_{\text{beam}}$ is the beam energy in the same frame.
The beam-energy-constrained mass is defined as
$\mbc = \sqrt{E_{\text{beam}}^2-(\sum_i\vec{p}_i)^2}$,
where $\vec{p}_i$ are the momenta of the $B$ decay products in the
center-of-mass frame. We retain $B$ candidates satisfying the conditions
$5.2 < \mbc < 5.3\ \gevcc$ and $|\de| < 0.2\ \gev$.

To improve the resolution on the invariant masses of mother
particles and $\de$, a mass-constrained fit is applied
to the $\piz$, $\eta$, $\eta'$, $h_c$, and $B$ candidates.
A fit with mass and vertex
constraints is applied to the $\ks$ and $\Lambda$ candidates.

In addition, the $\eta_{c2}(1D)$ daughter $\gamma$ energy is
required to be greater than $120\ \mev$ in the $B$ rest frame.
This requirement removes background from low-energy
photons. The signal efficiency of this requirement
is 100\%, because the $h_c \gamma$ invariant mass
is less than $3.7\ \gevcc$ for all excluded events.

To reduce continuum backgrounds, the ratio of the
Fox-Wolfram moments~\cite{Fox:1978vu} $F_2/F_0$ is
required to be less than 0.3.
For the two-body decays $\decaykp$ and $\decayks$, this requirement rejects
between 10\% to 44\% of background, depending on the $\eta_c$
decay channel, while the signal efficiency ranges from 94.4\% to 97.2\%.
For the three-body decays $\decaypimkp$ and $\decaypimks$, the
signal efficiency is in the range from 95.1\% to 97.3\%
and the fraction of the background rejected is (6-28)\%.

\section{Multivariate analysis and optimization of the selection requirements}

\subsection{General analysis strategy and data samples}

To improve the separation between the signal and background, we perform a
multivariate analysis followed by a global optimization
of the selection requirements. The first stages of the analysis are
performed individually for each $\eta_c$ decay channel.
They include the determination of two-dimensional $(\de,\mbc)$ resolution,
fit to the $(\de,\mbc)$ distribution, and the multivariate-analysis stage.
The global optimization of the selection requirements uses the results of all
initial stages as its input. The resolution is used to determine
the expected number of the signal events and the distribution of
the background in $(\de,\mbc)$ is used to determine the expected number
of the background events in the signal region.
The data selected using the resulting channel-dependent
criteria are merged into a single sample.

The experimental data are used for determination of the $(\de,\mbc)$
distribution, selection of the background samples for the neural network, and
final fit to the selected events.
During the development of the analysis procedure, the $\eta_{c2}(1D)$
region was excluded to avoid bias of the $\eta_{c2}(1D)$ significance.
The final fit described in section~\ref{sec:fit} was performed on MC
pseudoexperiments generated in accordance with the fit result without
the $\eta_{c2}(1D)$ mixed with the signal MC.
The excluded region is defined by
\begin{equation}
\sqrt{\left(\frac{\de}{\sigma_{\de}}\right)^2 +
\left(\frac{\mbc - m_B}{\sigma_{\mbc}}\right)^2} < 3,
\label{eq:blinded_region_de_mbc}
\end{equation}
where $\sigma_{\de} = 15\ \mev$ and $\sigma_{\mbc} = 2.5\ \mevcc$ are
the approximate resolutions in $\de$ and $\mbc$, respectively, and
\begin{equation}
3795 < M_{\eta_{c2}(1D)} < 3845\ \mevcc.
\label{eq:blinded_region_etac2}
\end{equation}
The requirement given by eq.~\eqref{eq:blinded_region_etac2} corresponds to
the $\eta_{c2}(1D)$ search region, chosen to be within $25\ \mevcc$
from the central value of $3820\ \mevcc$. The central value is chosen taking
into account the prediction that the difference of the $\eta_{c2}(1D)$
and $\psi_2(1D)$ masses is small.
After completion of the analysis procedure development,
this requirement is no longer used for the event selection.

The signal MC is used for the determination of the resolution and
the selection of the signal samples for the neural network.
The signal MC is generated using known information about the angular or
invariant-mass distributions of the decay products if this is possible;
otherwise, uniform distributions are assumed.
The multidimensional angular distribution is calculated for the decays
$\decaykp$ and $\decayks$
using the helicity formalism assuming a pure E1 transition between
the $\eta_{c2}(1D)$ and $h_c$. The $\eta_c$ decay resonant structure is also
taken into account if it is known. The distributions for the channels
$\etaccha{0}$, $\etaccha{1}$, $\etaccha{2}$, $\etaccha{3}$, and $\etaccha{4}$
are based on the results of a Dalitz plot analysis performed
in ref.~\cite{Lees:2014iua}. The contributions of intermediate $\phi$
resonances are taken into account for the channel $\etaccha{5}$ based on
the world-average branching fractions from ref.~\cite{Tanabashi:2018oca}.

\subsection{Resolution}

\begin{figure}
\centering
\includegraphics[width=7.0cm]{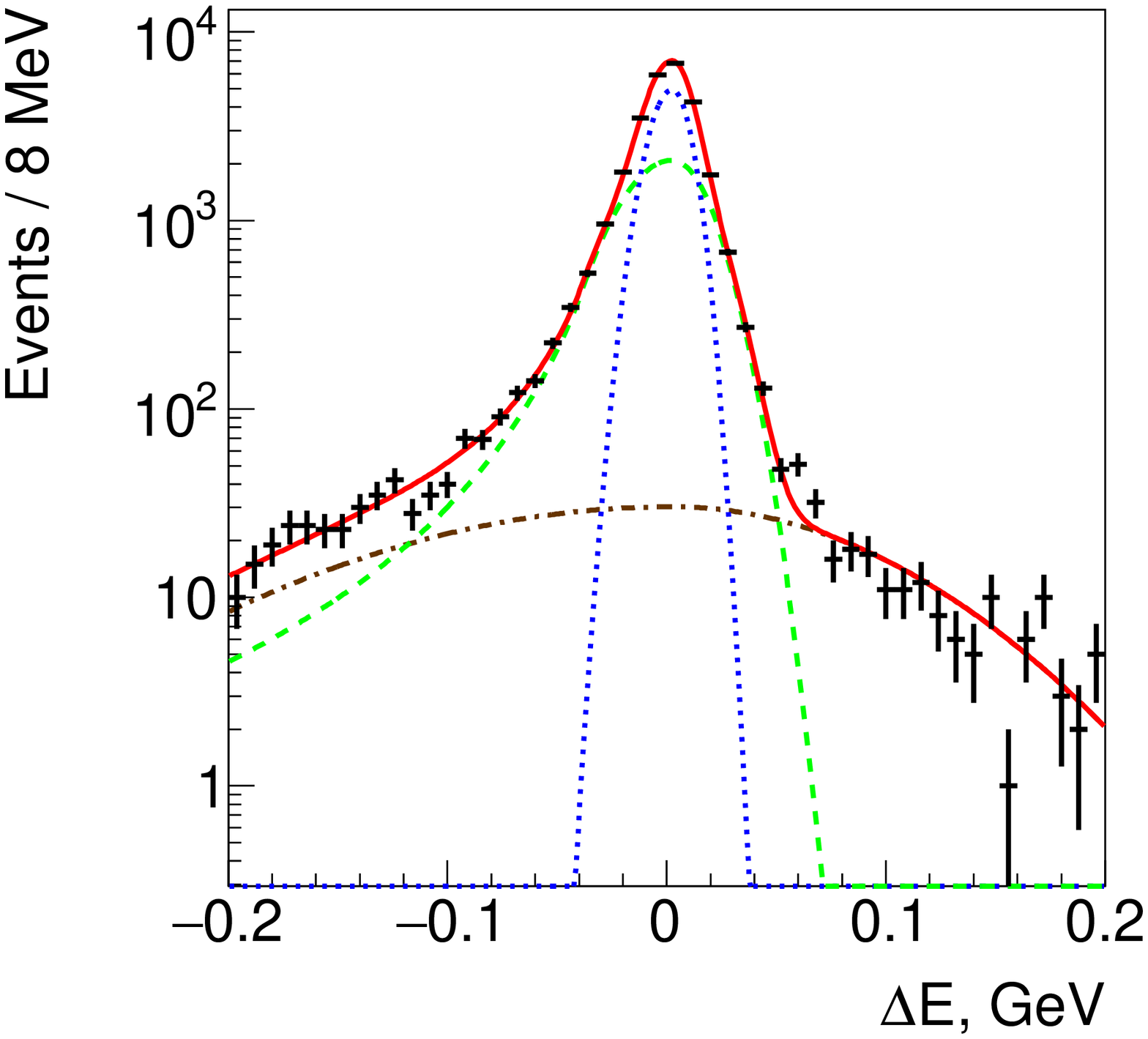}
\includegraphics[width=7.0cm]{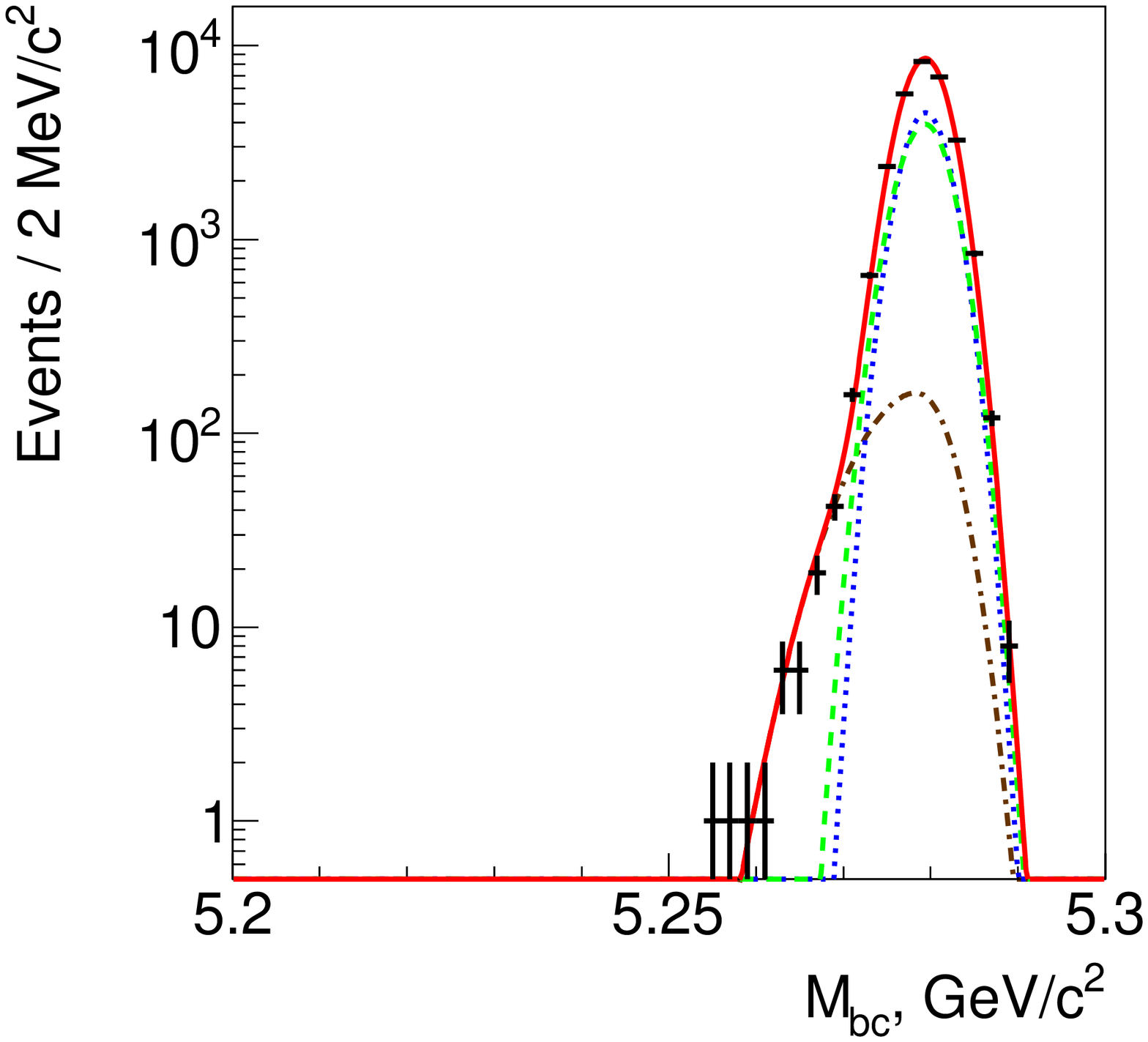} \\
\caption{Projections of the resolution fit results onto $\de$ and
$\mbc$ for the channel $\decaykp$ with $\etacdec{0}$.
The points with error bars are truth-matched signal MC,
the red solid curve is the fit result, the green
dashed curve is the Crystal Ball component, the blue dotted curve is the first
Gaussian component, and the brown dash-dotted curve is the second Gaussian
component.}
\label{fig:resolution}
\end{figure}

The first stages of the analysis procedure are the determination of the
$(\de,\mbc)$ resolution and fit to the $(\de,\mbc)$ distribution in data.
These two stages are performed exactly in the same way as in
ref.~\cite{Chilikin:2019wzy}.
The resolution is parameterized by the function
\begin{equation}
\begin{aligned}
S(\de, \mbc) = & N_{\text{CB}} F_{\text{CB}}(x_1) G_a^{(1 2)}(y_1)
+ N_\text{G1} G_a^{(2 1)}(x_2) G_a^{(2 2)}(y_2) \\
& + N_\text{G2} G_a^{(3 1)}(x_3) G_a^{(3 2)}(y_3), \\
\label{eq:signal_pdf}
\end{aligned}
\end{equation}
where $F_{\text{CB}}$ is an asymmetric Crystal Ball
function~\cite{Skwarnicki:1986xj},
$G_a^{(i j)}$ are asymmetric Gaussian functions,
$N_{\text{CB}}$, $N_\text{G1}$ and $N_\text{G2}$ are normalizations and
$x_i$ and $y_i$ ($i$ = 1, 2, 3) are rotated variables
that are given by
\begin{equation}
\begin{pmatrix}
x_i \\
y_i \\
\end{pmatrix}
=
\begin{pmatrix}
\cos\alpha_i & \sin\alpha_i \\
-\sin\alpha_i & \cos\alpha_i \\
\end{pmatrix}
\begin{pmatrix}
\de - (\de)_0 \\
\mbc - (\mbc)_0 \\
\end{pmatrix}.
\end{equation}
Here, ($(\de)_0$, $(\mbc)_0$) is the central point and
$\alpha_i$ is the rotation angle. The central point is
the same for all three terms in eq.~\eqref{eq:signal_pdf}.
The rotation eliminates the correlation of $\Delta E$
and $M_{bc}$, allowing the use of a Crystal Ball function for the
uncorrelated variable $x_1$.
The resolution is determined from a binned maximum likelihood fit
to signal MC events. Example resolution fit results (for the channel $\decaykp$
with $\etacdec{0}$) are shown in figure~\ref{fig:resolution}.

\subsection{Fit to the $(\de,\mbc)$ distribution}

\begin{figure}
\centering
\includegraphics[width=7.0cm]{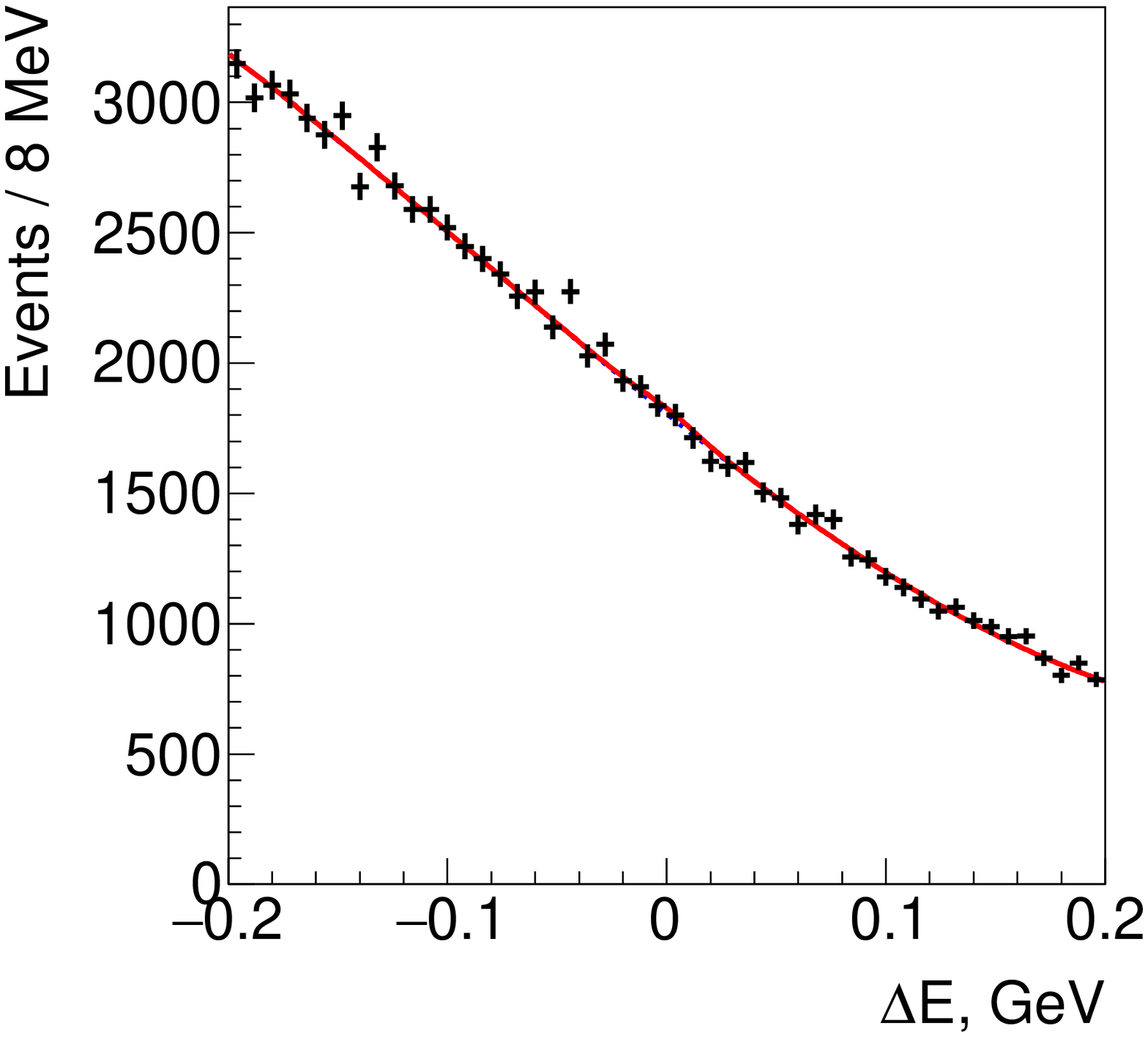}
\includegraphics[width=7.0cm]{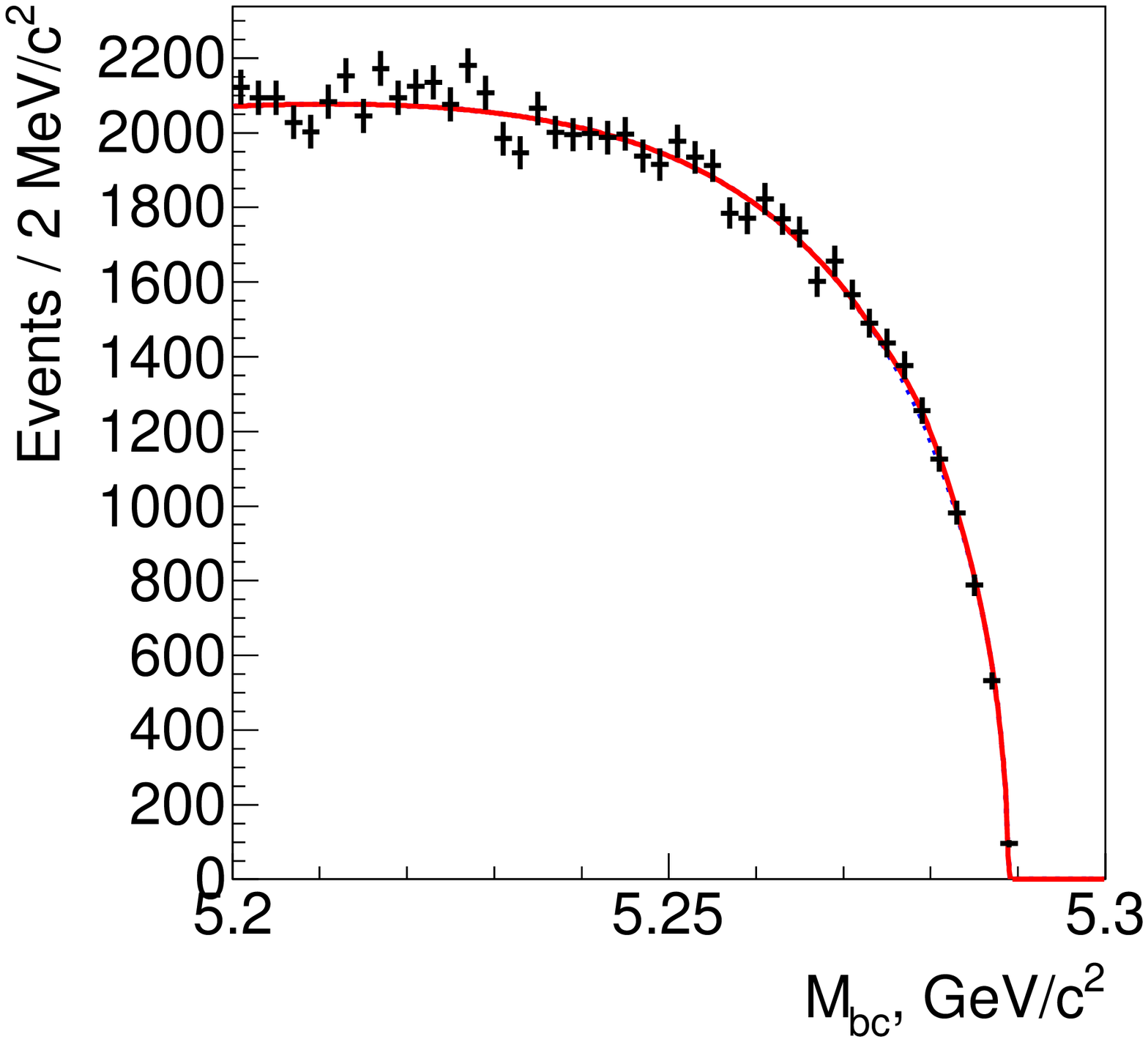} \\
\caption{Projections of the results of the fit to
the $(\de,\mbc)$ distribution onto $\de$
(with $\mbc > 5.272\ \gevcc$) and $\mbc$ (with $|\de| < 20\ \mev$)
for the channel $\decaykp$ with $\etacdec{0}$.
The points with error bars are data,
the red solid curve is the fit
result, and the blue dotted curve is the background. Since there is no
significant signal before the optimization of the selection requirements and
for the entire $\hccha{0}$ mass range, the two curves almost coincide.}
\label{fig:background}
\end{figure}

The $(\de,\mbc)$ distribution is fitted in order to determine the expected
number of the background events in the signal region.
The distribution is fitted using the function
\begin{equation}
N_S S(\de,\mbc) + B(\de,\mbc),
\end{equation}
where $N_S$ is the number of signal events and $B$ is the background density
function that is given by
\begin{equation}
B(\de,\mbc) = \sqrt{m_0 - \mbc} \exp[-a (m_0 - \mbc)] P_3(\de,\mbc),
\label{eq:background_density}
\end{equation}
where $m_0$ is the threshold mass,
$a$ is a rate parameter and $P_3$ is a two-dimensional third-order polynomial.
Example $(\de,\mbc)$ fit results (for the channel $\decaykp$ with $\etacdec{0}$)
are shown in figure~\ref{fig:background}.

\subsection{Multivariate analysis}
\label{sec:mva}

To improve the separation of signal and background events,
a multivariate analysis is performed for each individual $\eta_c$ decay
channel. As in ref.~\cite{Chilikin:2019wzy}, the multivariate analysis
is performed using the multilayer perceptron (MLP) neural network implemented
in the {\sc tmva} library~\cite{tmva}. However, the details of the procedure
depend on the particular $B$ decay mode and, consequently, differ from
ref.~\cite{Chilikin:2019wzy}.

The following variables are always included in the neural network:
the angle between the thrust axes of the $B$ candidate and the remaining
particles in the event, the angle between the thrust axes of all tracks
and all photons in the event, the ratio of the Fox-Wolfram moments
$F_2/F_0$, the $B$ production angle ($\Upsilon(4S)$ helicity angle),
the quality of the vertex fit of all $B$ daughter tracks, $\ks$,
and $\lam$ to a common vertex,
the $h_c$ and $\eta_c$ masses, and the maximal $\piz$ likelihoods
for the $\eta_{c2}(1D)$ and $h_c$ daughter photons combined with
any other photon in the event. The
$\piz$ likelihood is based on the energy and the polar angle
of the transition-photon candidate in the laboratory frame
and the $\piz$ invariant mass.
Note that the ratio of the Fox-Wolfram moments is already
required to be less than 0.3 at the event-selection stage. This requirement
does not significantly modify the final selection results, because
the rejection of background with $F_2/F_0>0.3$ can also be performed by the MLP.
However, it lowers the fraction of the background that needs to be rejected
by the MLP, which is helpful to reduce overtraining.

\begin{figure}
\centering
\includegraphics[width=7.0cm]{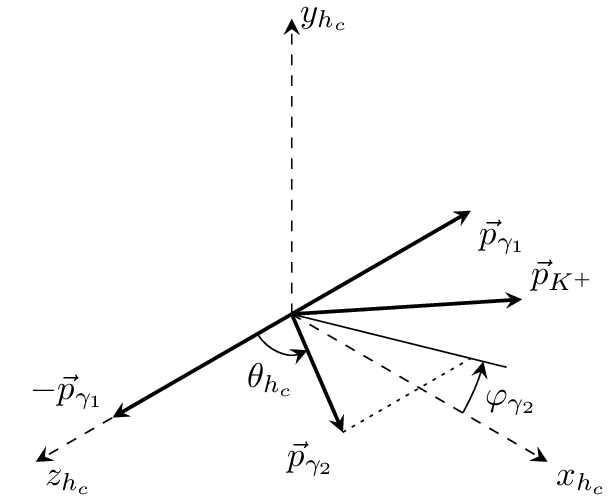} \\
\caption{Definition of the azimuthal angle $\varphi_{\gamma_2}$ of
the $h_c$ daughter photon (in the $h_c$ rest frame). The $\eta_{c2}(1D)$
daughter photon is denoted as $\gamma_1$.}
\label{fig:angle_definition}
\end{figure}

For the decays $\decaykp$ and $\decayks$, the MLP also includes
the $\eta_{c2}(1D)$ helicity angle and the $h_c$ daughter-photon
azimuthal angle for every $\eta_c$ channel. 
The $\eta_{c2}(1D)$ helicity angle is defined as the angle between
$-\vec{p}_{K}$ and $\vec{p}_{h_c}$, where $\vec{p}_{K}$ and $\vec{p}_{h_c}$
are the momenta of $K$ and $h_c$ in the $\eta_{c2}(1D)$ rest frame,
respectively.
The azimuthal angle of the $h_c$ daughter photon is defined as the
angle between the planes of the $h_c$ and $\eta_{c2}(1D)$
daughter-photon momenta and the momenta of the $\eta_{c2}(1D)$
daughter photon and $\kp$ or $\ks$ in the
$h_c$ rest frame. The definition is shown in figure~\ref{fig:angle_definition}.

For the channels $\etacdec{0}$, $\etacdec{1}$, and $\etacdec{2}$, two
invariant masses of the $\eta_c$ daughter particle pairs
(both $K \pi$ combinations) are included in the neural network.

The following particle identification variables are included into
the neural network if there are corresponding charged particles
in the final state: the minimum likelihood ratio $R_{K/\pi}$ of
the $\eta_c$ daughter kaons, the minimum of the two likelihood ratios
$R_{p/K}$, $R_{p/\pi}$ of the $\eta_c$ daughter protons, and
$R_{K/\pi}$ for the daughter $\kp$ of the $B$ meson (for the decays
$\decaykp$ and $\decaypimkp$).

If there is a $\piz$ or $\eta$ in the final state,
four additional variables are included:
the $\piz$ ($\eta$) mass,
the minimal energy of the $\piz$ ($\eta$) daughter photons in the
laboratory frame, and the number of $\piz$ candidates that include a
$\piz$ ($\eta$) daughter photon as one of their daughters (for each of
the $\piz$ ($\eta$) daughter photons).
If the $\eta_c$ has an $\eta'$ daughter, then the mass of the $\eta'$
candidate is also included in the neural network.

The training and testing signal samples are taken from
the signal MC.
The background sample is taken from a two-dimensional $(\de,\mbc)$ sideband
defined as all selected events outside the signal region defined by
eq.~\ref{eq:blinded_region_de_mbc}. The background sample is divided into
training and testing samples.

\subsection{Global optimization of the selection requirements}
\label{sec:global_optimization}

The global selection-requirement optimization is performed by maximizing the
value
\begin{equation}
F_\text{opt} = \frac{\sum\limits_i N_\text{sig}^{(i)}}
{\displaystyle{\frac{a}{2}} + \sqrt{\sum\limits_i N_\text{bg}^{(i)}}},
\label{eq:optfun}
\end{equation}
where $i$ is the channel index, $N_\text{sig}^{(i)}$
and $N_\text{bg}^{(i)}$ are the expected numbers of the signal and 
background events for the $i$-th channel in the signal region, respectively,
and $a=3$ is the target significance.
This optimization method is based on ref.~\cite{Punzi:2003bu}.

The signal region is defined as
\begin{equation}
\left(\frac{\de}{R_{\de}^{(i)}}\right)^2 +
\left(\frac{\mbc - m_{B}}{R_{\mbc}^{(i)}}\right)^2 < 1.
\end{equation}
where $R_{\de}^{(i)}$ and $R_{\mbc}^{(i)}$
are the half-axes of the signal region ellipse.
The parameters determined by the optimization are
$R_{\de}^{(i)}$, $R_{\mbc}^{(i)}$,
and the minimal value of the MLP output ($v_0^{(i)}$)
for each channel.

The expected number of signal events for $\decaykp$ is calculated as
\begin{equation}
\begin{aligned}
N_\text{sig}^{(i)} = & 2 N_{\Upsilon(4S)} \br(\Upsilon(4S)\to\Bp\Bm)
\br(\decaykp) \br(\eta_{c2}(1D) \to h_c \gamma) \\
& \times \br(h_c \to \eta_c \gamma)
\br(\eta_c \to i) \epsilon_\text{SR}^{(i)}
\epsilon_S^{(i)}(v_0^{(i)}),
\end{aligned}
\end{equation}
where $N_{\Upsilon(4S)}$ is the number of $\Upsilon(4S)$ events,
$\br(\eta_c \to i)$ is the branching fraction of the $\eta_c$ decay
to its $i$-th decay channel,
$\epsilon_\text{SR}^{(i)}$ is the reconstruction efficiency for the
specific signal region $\text{SR}$
before the requirement ($v > v_0^{(i)}$)
on the MLP output variable $v$ for the signal events, and
$\epsilon_S^{(i)}(v_0^{(i)})$ is the efficiency of the MLP-output requirement.
The number of $\Upsilon(4S)$ events is assumed to be equal to the number of
$B \bar{B}$ pairs; the branching fraction $\br(\Upsilon(4S)\to\Bp\Bm)$
is calculated under the same assumption in ref.~\cite{Tanabashi:2018oca}.
The signal-region-dependent reconstruction efficiency is calculated as
\begin{equation}
\epsilon_\text{SR}^{(i)} = \epsilon_R^{(i)}
\int\limits_\text{SR} S_i(\de,\mbc) d \de d \mbc,
\end{equation}
where $\epsilon_R^{(i)}$ is the reconstruction efficiency,
and $S_i$ is the signal probability density function
for $i$-th $\eta_c$
decay channel (the integral of $S_i$ over the signal region is the
efficiency of the signal region selection).
The unknown branching-fraction product
$\br(\decaykp) \times \br(\eta_{c2}(1D) \to h_c \gamma)$
can be set to an arbitrary value because the maximum of
$F_\text{opt}$ does not depend on it. The expected number of signal events
for $\decayks$ is calculated in a similar manner.

The expected number of background events is calculated as
\begin{equation}
N_\text{bg}^{(i)} = \epsilon_B^{(i)}(v_0^{(i)})
\frac{N_{\eta_{c2}(1D)\text{ region}}}{N_\text{full}}
\int\limits_\text{SR} B_i(\de,\mbc) d \de d\mbc,
\end{equation}
where $\epsilon_B^{(i)}(v_0^{(i)})$ is the efficiency of the MLP output
requirement for the background events,
$N_{\eta_{c2}(1D)\text{ region}}$ is the number of
background events in the $\eta_{c2}(1D)$ region defined by
eq.~\eqref{eq:blinded_region_etac2}, $N_\text{full}$ is the full number
of background events, and
$B_i$ is the background density function.

The results are shown
in table~\ref{tab:etac2kp_optimization} for the decay $\decaykp$,
in table~\ref{tab:etac2ks_optimization} for the decay $\decayks$,
in table~\ref{tab:etac2k_optimization} for the combination
of $\decaykp$ and $\decayks$,
in table~\ref{tab:etac2pimkp_optimization} for the decay $\decaypimkp$,
and in table~\ref{tab:etac2pimks_optimization} for the decay $\decaypimks$.
To combine the decays $\decaykp$ and $\decayks$,
a separate optimization that includes all combinations of $B$ and $\eta_c$
channels is performed. To estimate the expected number of signal events
in the combined sample, the partial widths of the decays $\decaykp$ and
$\Bz \to \eta_{c2}(1D) K^0$ are assumed to be the same.

\begin{table}
\caption{Results of the optimization of the selection requirements
for the channel $\decaykp$.
The expected number of signal events is calculated assuming
$\br(\decaykp) \times \br(\eta_{c2}(1D) \to h_c \gamma) = 1.0 \times 10^{-5}$.
The efficiencies and expected numbers of signal and background events
are calculated for the training sample.
The signal-region half-axes $R_{\de}^{(i)}$ ($R_{\mbc}^{(i)}$)
are in $\mev$ ($\mevcc$); all other values are dimensionless.}
\centering
\begin{tabular}{c|c|c|c|c|c|c|c}
\hline\hline
\multirow{2}{*}{Channel} & \multicolumn{2}{|c|}{Parameters} & \multicolumn{3}{|c|}{Efficiency} & \multicolumn{2}{|c}{Events} \\
\cline{2-8}
 & $R_{\de}^{(i)}$ & $R_{\mbc}^{(i)}$ & $\epsilon_\text{SR}^{(i)}$ & $\epsilon_S^{(i)}(v_0^{(i)})$ & $\epsilon_B^{(i)}(v_0^{(i)})$ & $N_\text{sig}^{(i)}$ & $N_\text{bg}^{(i)}$ \\
\hline
$\optcha{0}$ & 23.3 & 4.85 & 4.29\% & 51.3\% & 2.22\% & 2.16 & 9.82 \\
$\optcha{1}$ & 18.4 & 4.11 & 2.76\% & 32.8\% & 0.37\% & 0.44 & 2.82 \\
$\optcha{2}$ & 18.0 & 3.80 & 0.93\% & 21.5\% & 0.20\% & 0.02 & 0.17 \\
$\optcha{3}$ & 21.5 & 4.70 & 2.93\% & 15.2\% & 0.07\% & 0.05 & 0.26 \\
$\optcha{5}$ & 20.0 & 4.33 & 3.47\% & 42.1\% & 1.97\% & 0.09 & 0.49 \\
$\optcha{9}$ & 21.3 & 4.64 & 2.02\% & 24.1\% & 0.13\% & 0.09 & 0.47 \\
$\optcha{13}$ & 30.1 & 5.61 & 12.53\% & 66.4\% & 4.50\% & 0.51 & 1.52 \\
$\optcha{14}$ & 19.5 & 4.24 & 3.36\% & 21.1\% & 0.16\% & 0.10 & 0.62 \\
$\optcha{15}$ & 18.0 & 3.99 & 4.13\% & 21.1\% & 0.46\% & 0.19 & 1.18 \\
$\optcha{16}$ & 30.4 & 5.67 & 2.71\% & 59.9\% & 3.16\% & 0.07 & 0.22 \\
\hline\hline
\end{tabular}
\label{tab:etac2kp_optimization}
\end{table}

\begin{table}
\caption{Results of the optimization of the selection requirements
for the channel $\decayks$.
The expected number of signal events is calculated assuming
$\br(\decaykp) \times \br(\eta_{c2}(1D) \to h_c \gamma) = 1.0 \times 10^{-5}$
and equal partial widths of the decays $\decaykp$ and
$\Bz \to \eta_{c2}(1D) K^0$.
The efficiencies and expected numbers of signal and background events
are calculated for the training sample.
The signal-region half-axes $R_{\de}^{(i)}$ ($R_{\mbc}^{(i)}$)
are in $\mev$ ($\mevcc$); all other values are dimensionless.}
\centering
\begin{tabular}{c|c|c|c|c|c|c|c}
\hline\hline
\multirow{2}{*}{Channel} & \multicolumn{2}{|c|}{Parameters} & \multicolumn{3}{|c|}{Efficiency} & \multicolumn{2}{|c}{Events} \\
\cline{2-8}
 & $R_{\de}^{(i)}$ & $R_{\mbc}^{(i)}$ & $\epsilon_\text{SR}^{(i)}$ & $\epsilon_S^{(i)}(v_0^{(i)})$ & $\epsilon_B^{(i)}(v_0^{(i)})$ & $N_\text{sig}^{(i)}$ & $N_\text{bg}^{(i)}$ \\
\hline
$\optcha{0}$ & 24.5 & 5.06 & 2.75\% & 61.2\% & 4.27\% & 0.73 & 4.88 \\
$\optcha{1}$ & 19.2 & 4.12 & 1.77\% & 45.8\% & 1.09\% & 0.17 & 1.85 \\
$\optcha{2}$ & 17.0 & 3.51 & 0.51\% & 33.5\% & 0.61\% & 0.01 & 0.12 \\
$\optcha{3}$ & 20.0 & 4.43 & 1.72\% & 25.7\% & 0.29\% & 0.02 & 0.21 \\
$\optcha{5}$ & 22.5 & 4.73 & 2.24\% & 49.6\% & 3.06\% & 0.03 & 0.22 \\
$\optcha{9}$ & 21.9 & 4.64 & 1.21\% & 32.7\% & 0.32\% & 0.03 & 0.29 \\
$\optcha{13}$ & 34.7 & 6.06 & 8.23\% & 70.6\% & 6.73\% & 0.16 & 0.65 \\
$\optcha{14}$ & 19.5 & 4.29 & 2.08\% & 26.8\% & 0.34\% & 0.04 & 0.36 \\
$\optcha{15}$ & 19.9 & 4.27 & 2.71\% & 30.1\% & 0.92\% & 0.08 & 0.74 \\
$\optcha{16}$ & 34.0 & 6.06 & 1.71\% & 65.1\% & 4.66\% & 0.02 & 0.09 \\
\hline\hline
\end{tabular}
\label{tab:etac2ks_optimization}
\end{table}

\begin{table}
\caption{Results of the optimization of the selection requirements
for the combination of the channels $\decaykp$ and $\decayks$.
The expected number of signal events is calculated assuming
$\br(\decaykp) \times \br(\eta_{c2}(1D) \to h_c \gamma) = 1.0 \times 10^{-5}$
and equal partial widths of the decays $\decaykp$ and
$\Bz \to \eta_{c2}(1D) K^0$.
The efficiencies and expected numbers of signal and background events
are calculated for the training sample.
The signal-region half-axes $R_{\de}^{(i)}$ ($R_{\mbc}^{(i)}$)
are in $\mev$ ($\mevcc$); all other values are dimensionless.}
\centering
\begin{tabular}{c|c|c|c|c|c|c|c}
\hline\hline
\multirow{2}{*}{Channel} & \multicolumn{2}{|c|}{Parameters} & \multicolumn{3}{|c|}{Efficiency} & \multicolumn{2}{|c}{Events} \\
\cline{2-8}
 & $R_{\de}^{(i)}$ & $R_{\mbc}^{(i)}$ & $\epsilon_\text{SR}^{(i)}$ & $\epsilon_S^{(i)}(v_0^{(i)})$ & $\epsilon_B^{(i)}(v_0^{(i)})$ & $N_\text{sig}^{(i)}$ & $N_\text{bg}^{(i)}$ \\
\hline
\multicolumn{8}{c}{$\decaykp$} \\
\hline
$\optcha{0}$ & 24.3 & 4.93 & 4.37\% & 45.3\% & 1.55\% & 1.95 & 7.27 \\
$\optcha{1}$ & 17.0 & 3.90 & 2.58\% & 32.8\% & 0.37\% & 0.41 & 2.48 \\
$\optcha{2}$ & 16.8 & 3.52 & 0.86\% & 21.5\% & 0.20\% & 0.02 & 0.15 \\
$\optcha{3}$ & 20.4 & 4.45 & 2.80\% & 15.2\% & 0.07\% & 0.05 & 0.23 \\
$\optcha{5}$ & 20.8 & 4.49 & 3.58\% & 34.3\% & 1.25\% & 0.07 & 0.33 \\
$\optcha{9}$ & 20.2 & 4.47 & 1.94\% & 24.1\% & 0.13\% & 0.09 & 0.43 \\
$\optcha{13}$ & 28.4 & 5.48 & 12.27\% & 66.4\% & 4.50\% & 0.50 & 1.40 \\
$\optcha{14}$ & 18.4 & 4.04 & 3.19\% & 20.8\% & 0.16\% & 0.10 & 0.54 \\
$\optcha{15}$ & 16.7 & 3.83 & 3.88\% & 20.9\% & 0.45\% & 0.17 & 1.04 \\
$\optcha{16}$ & 28.8 & 5.55 & 2.66\% & 59.9\% & 3.16\% & 0.07 & 0.20 \\
\hline
\multicolumn{8}{c}{$\decayks$} \\
\hline
$\optcha{0}$ & 24.8 & 4.92 & 2.73\% & 44.9\% & 1.71\% & 0.53 & 1.91 \\
$\optcha{1}$ & 17.6 & 3.87 & 1.64\% & 28.3\% & 0.40\% & 0.10 & 0.59 \\
$\optcha{2}$ & 16.1 & 3.43 & 0.49\% & 9.0\% & 0.09\% & 0.00 & 0.02 \\
$\optcha{3}$ & 14.0 & 3.17 & 1.17\% & 25.7\% & 0.29\% & 0.01 & 0.11 \\
$\optcha{5}$ & 19.2 & 4.18 & 2.00\% & 41.5\% & 2.04\% & 0.02 & 0.11 \\
$\optcha{9}$ & 19.8 & 4.28 & 1.12\% & 24.7\% & 0.16\% & 0.02 & 0.12 \\
$\optcha{13}$ & 32.7 & 5.57 & 8.00\% & 65.4\% & 4.57\% & 0.14 & 0.38 \\
$\optcha{14}$ & 16.0 & 3.62 & 1.72\% & 18.8\% & 0.18\% & 0.02 & 0.13 \\
$\optcha{15}$ & 16.7 & 3.74 & 2.33\% & 22.8\% & 0.52\% & 0.05 & 0.30 \\
$\optcha{16}$ & 35.5 & 6.15 & 1.73\% & 54.9\% & 1.78\% & 0.02 & 0.04 \\
\hline\hline
\end{tabular}
\label{tab:etac2k_optimization}
\end{table}

\begin{table}
\caption{Results of the optimization of the selection requirements
for the channel $\decaypimkp$.
The expected number of signal events is calculated assuming
$\br(\decaypimkp) \times \br(\eta_{c2}(1D) \to h_c \gamma) = 1.0 \times 10^{-5}$.
The efficiencies and expected numbers of signal and background events
are calculated for the training sample.
The signal-region half-axes $R_{\de}^{(i)}$ ($R_{\mbc}^{(i)}$)
are in $\mev$ ($\mevcc$); all other values are dimensionless.}
\centering
\begin{tabular}{c|c|c|c|c|c|c|c}
\hline\hline
\multirow{2}{*}{Channel} & \multicolumn{2}{|c|}{Parameters} & \multicolumn{3}{|c|}{Efficiency} & \multicolumn{2}{|c}{Events} \\
\cline{2-8}
 & $R_{\de}^{(i)}$ & $R_{\mbc}^{(i)}$ & $\epsilon_\text{SR}^{(i)}$ & $\epsilon_S^{(i)}(v_0^{(i)})$ & $\epsilon_B^{(i)}(v_0^{(i)})$ & $N_\text{sig}^{(i)}$ & $N_\text{bg}^{(i)}$ \\
\hline
$\optcha{0}$ & 20.1 & 4.54 & 2.91\% & 40.3\% & 1.45\% & 1.09 & 33.26 \\
$\optcha{1}$ & 15.6 & 3.64 & 1.83\% & 24.9\% & 0.26\% & 0.21 & 9.29 \\
$\optcha{2}$ & 12.1 & 2.87 & 0.43\% & 20.3\% & 0.21\% & 0.01 & 0.56 \\
$\optcha{3}$ & 13.6 & 3.32 & 1.44\% & 16.7\% & 0.11\% & 0.02 & 1.22 \\
$\optcha{5}$ & 16.8 & 3.92 & 2.22\% & 23.2\% & 0.61\% & 0.03 & 1.14 \\
$\optcha{9}$ & 17.3 & 3.92 & 1.18\% & 20.7\% & 0.10\% & 0.04 & 1.75 \\
$\optcha{13}$ & 25.0 & 5.23 & 9.01\% & 61.9\% & 5.24\% & 0.32 & 7.10 \\
$\optcha{14}$ & 14.4 & 3.48 & 1.97\% & 16.8\% & 0.14\% & 0.04 & 2.09 \\
$\optcha{15}$ & 16.3 & 3.81 & 2.74\% & 9.8\% & 0.12\% & 0.05 & 2.23 \\
$\optcha{16}$ & 24.8 & 5.23 & 1.82\% & 54.9\% & 3.48\% & 0.04 & 0.94 \\
\hline\hline
\end{tabular}
\label{tab:etac2pimkp_optimization}
\end{table}

\begin{table}
\caption{Results of the optimization of the selection requirements
for the channel $\decaypimks$.
The expected number of signal events is calculated assuming
$\br(\decaypimkp) \times \br(\eta_{c2}(1D) \to h_c \gamma) =
1.0 \times 10^{-5}$ and equal partial widths of the decays $\decaypimkp$ and
$\Bp \to \eta_{c2}(1D) \pip K^0$.
The efficiencies and expected numbers of signal and background events
are calculated for the training sample.
The signal-region half-axes $R_{\de}^{(i)}$ ($R_{\mbc}^{(i)}$)
are in $\mev$ ($\mevcc$); all other values are dimensionless.}
\centering
\begin{tabular}{c|c|c|c|c|c|c|c}
\hline\hline
\multirow{2}{*}{Channel} & \multicolumn{2}{|c|}{Parameters} & \multicolumn{3}{|c|}{Efficiency} & \multicolumn{2}{|c}{Events} \\
\cline{2-8}
 & $R_{\de}^{(i)}$ & $R_{\mbc}^{(i)}$ & $\epsilon_\text{SR}^{(i)}$ & $\epsilon_S^{(i)}(v_0^{(i)})$ & $\epsilon_B^{(i)}(v_0^{(i)})$ & $N_\text{sig}^{(i)}$ & $N_\text{bg}^{(i)}$ \\
\hline
$\optcha{0}$ & 21.6 & 4.69 & 1.64\% & 40.0\% & 1.64\% & 0.299 & 16.82 \\
$\optcha{1}$ & 15.6 & 3.57 & 0.94\% & 22.1\% & 0.33\% & 0.047 & 4.29 \\
$\optcha{2}$ & 12.4 & 2.82 & 0.24\% & 11.1\% & 0.14\% & 0.002 & 0.17 \\
$\optcha{3}$ & 15.8 & 3.67 & 0.90\% & 16.4\% & 0.12\% & 0.007 & 0.66 \\
$\optcha{5}$ & 17.9 & 4.11 & 1.24\% & 27.3\% & 0.85\% & 0.009 & 0.67 \\
$\optcha{9}$ & 12.9 & 3.01 & 0.45\% & 21.3\% & 0.21\% & 0.008 & 0.85 \\
$\optcha{13}$ & 27.4 & 5.39 & 5.06\% & 61.3\% & 5.40\% & 0.089 & 3.61 \\
$\optcha{14}$ & 16.1 & 3.67 & 1.14\% & 13.3\% & 0.11\% & 0.010 & 0.88 \\
$\optcha{15}$ & 16.3 & 3.88 & 1.46\% & 11.9\% & 0.17\% & 0.017 & 1.39 \\
$\optcha{16}$ & 26.1 & 5.36 & 0.99\% & 50.2\% & 3.42\% & 0.010 & 0.44 \\
\hline\hline
\end{tabular}
\label{tab:etac2pimks_optimization}
\end{table}

\section{Fit to the data}
\label{sec:fit}

\subsection{Fit results in the default model}

After the global optimization of the selection requirements, the selected
events are merged into a single data sample.
The best-candidate selection is performed for each $\eta_c$ channel separately
by using the maximal MLP output value;
the selection procedure is the same as in
ref.~\cite{Chilikin:2019wzy}. The fraction of removed candidates is
10\% to 23\%, depending on the $\eta_c$ channel, for the two-body decays
$\decaykp$ and $\decayks$; for the three-body decays $\decaypimkp$ and
$\decaypimks$, the fraction is (21-42)\%.
Multiple candidates originating from different $\eta_c$ decay channels are
allowed, however, no events with multiple candidates are observed in the
signal region for any of the signal $B$ decays.

We perform an extended unbinned maximum likelihood fit to the data in
the signal region.
The $\eta_{c2}(1D)$ is represented by the Breit-Wigner amplitude:
\begin{equation}
A_{\eta_{c2}(1D)}(M_{\eta_{c2}(1D)}) =
\frac{1}{m_{\eta_{c2}(1D)}^2 - M_{\eta_{c2}(1D)}^2 -
i M_{\eta_{c2}(1D)} \Gamma_{\eta_{c2}(1D)}},
\label{eq:breit_wigner}
\end{equation}
where $\Gamma_{\eta_{c2}(1D)}$ is the $\eta_{c2}(1D)$ width.
Since the $\eta_{c2}(1D)$ is expected to be narrower than the
resolution, it is sufficient to use the constant-width parameterization
given by eq.~\eqref{eq:breit_wigner}.
The $M_{h_c \gamma}$ distribution is fitted to the function
\begin{equation}
S(M) = \left(N_{\eta_{c2}(1D)} |A_{\eta_{c2}(1D)}(M)|^2\right)
\otimes R_{\eta_{c2}(1D)}(M) + P_2(M),
\end{equation}
where $N_{\eta_{c2}(1D)}$ is the number of signal events,
$R_{\eta_{c2}(1D)}(M)$ is the $\eta_{c2}(1D)$ mass resolution that is
determined from MC and parameterized as a sum of two asymmetric Gaussians,
and $P_2$ is a second-order polynomial representing the
background shape.
For the channel $\decayks$, that has the lowest number of events,
the default order of the background polynomial is reduced to 1.
The $\eta_{c2}(1D)$ width is fixed at $500\ \kev$.
The chosen default width value is approximately equal to
the sum of individual partial widths predicted in ref.~\cite{Eichten:2002qv}:
$\Gamma(\eta_{c2}(1D) \to h_c \gamma) + 
\Gamma(\eta_{c2}(1D) \to g g) +
\Gamma(\eta_{c2}(1D) \to \eta_c \pi \pi) = 458\ \kev$.
Another prediction of the $\eta_{c2}(1D)$ width was made in
ref.~\cite{Fan:2009cj}; it is estimated to be within the range
from $660$ to $810\ \kev$. The variation of the $\eta_{c2}(1D)$ width
is considered as a source of systematic uncertainty.

\begin{figure}
\centering
\includegraphics[width=7cm]{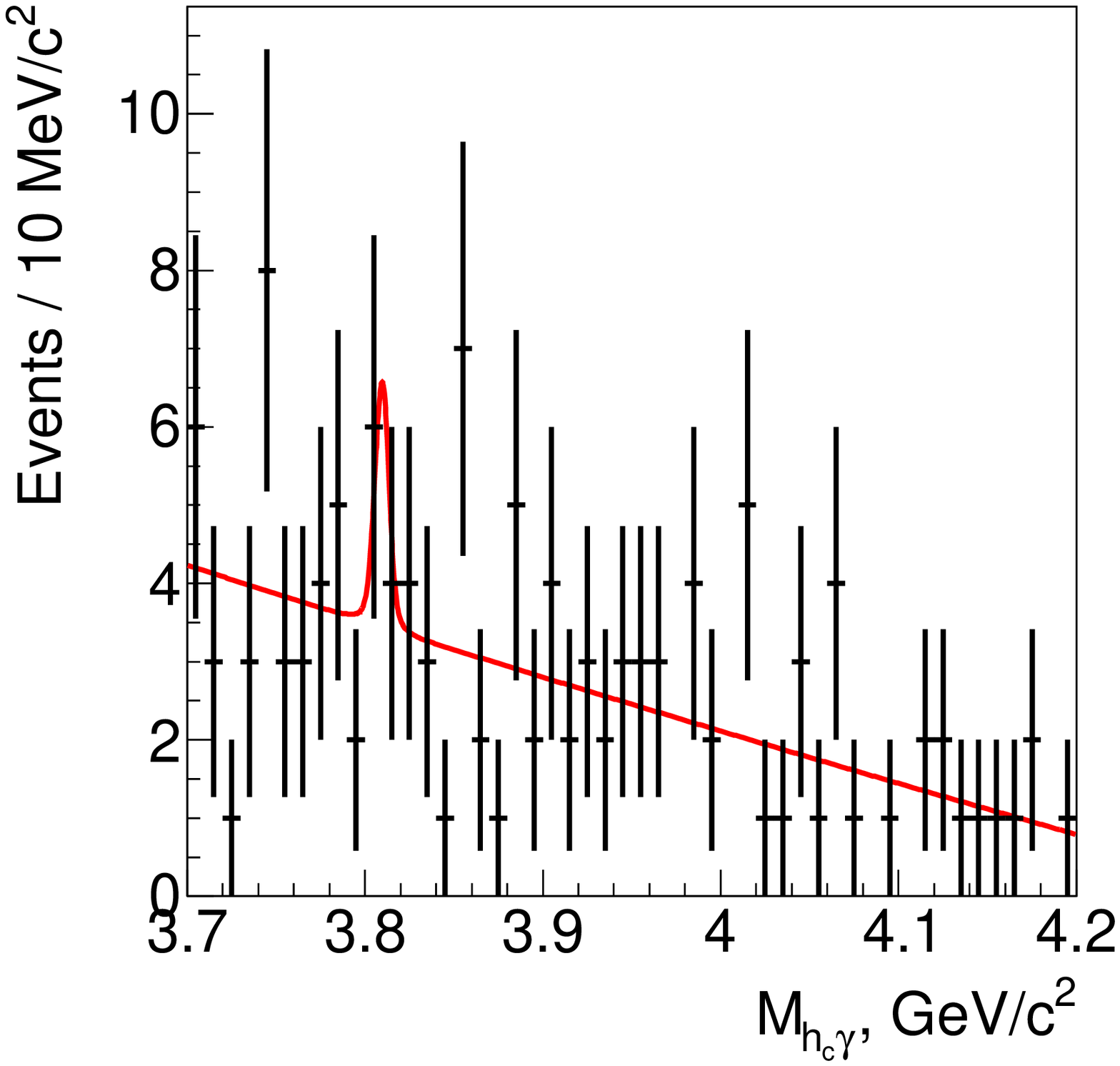}
\includegraphics[width=7cm]{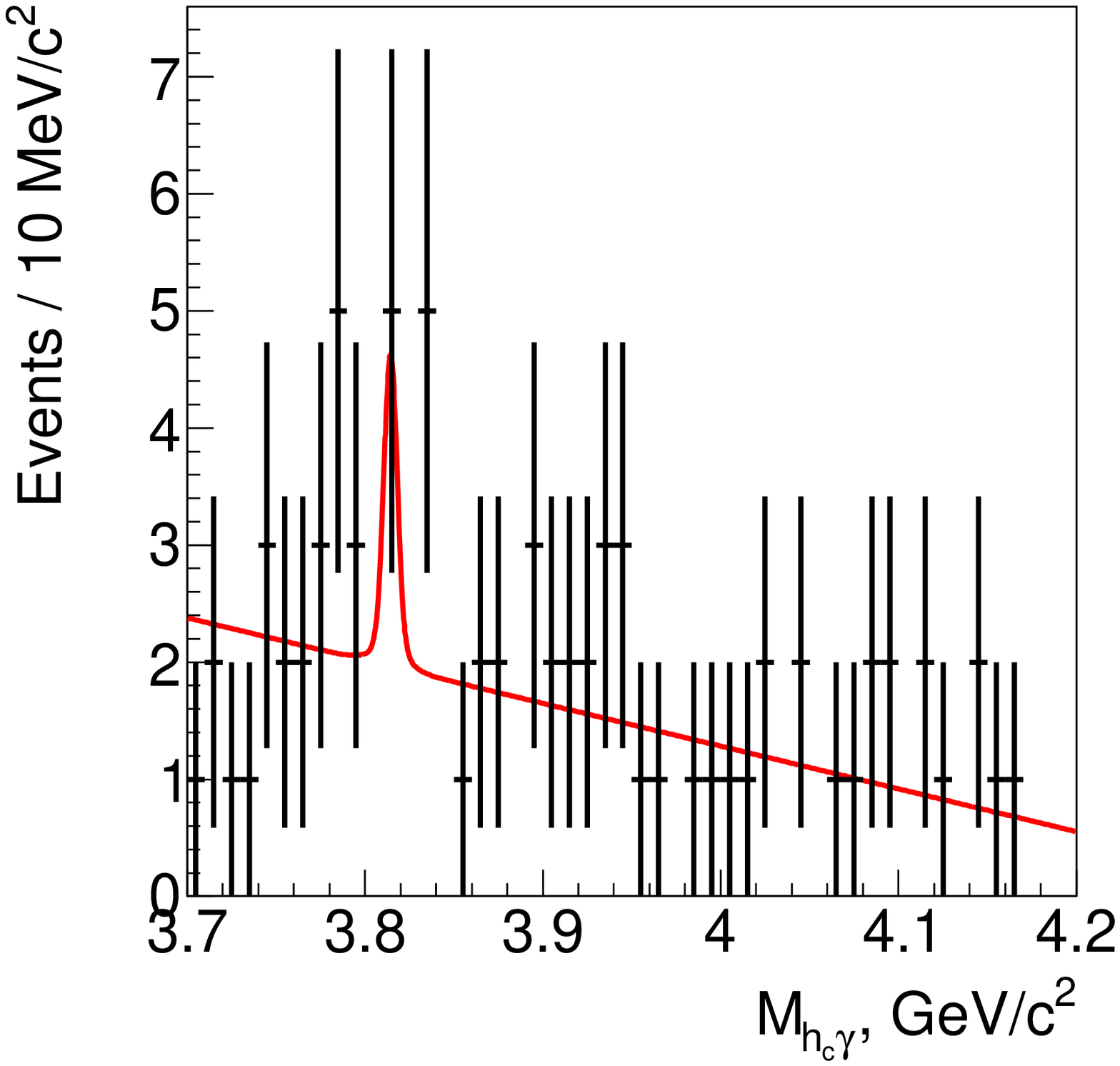} \\
\caption{Fit results for the channels $\decaykp$ (left) and $\decayks$ (right).}
\label{fig:etac2kp_fit_data}
\end{figure}

\begin{figure}
\centering
\includegraphics[width=7cm]{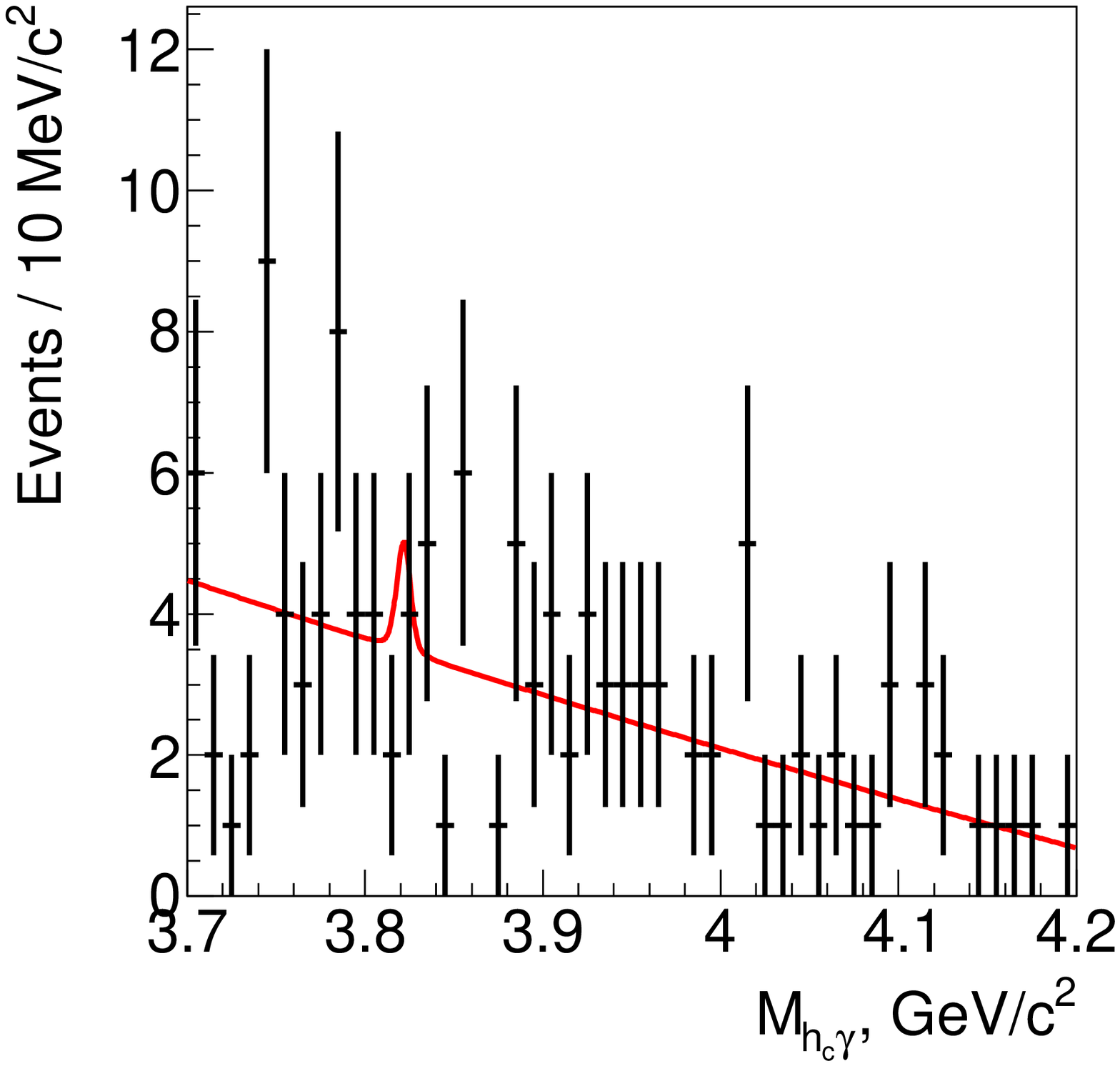} \\
\caption{Fit results for the combination of channels $\decaykp$ and $\decayks$.}
\label{fig:etac2k_fit_data}
\end{figure}

\begin{figure}
\centering
\includegraphics[width=7cm]{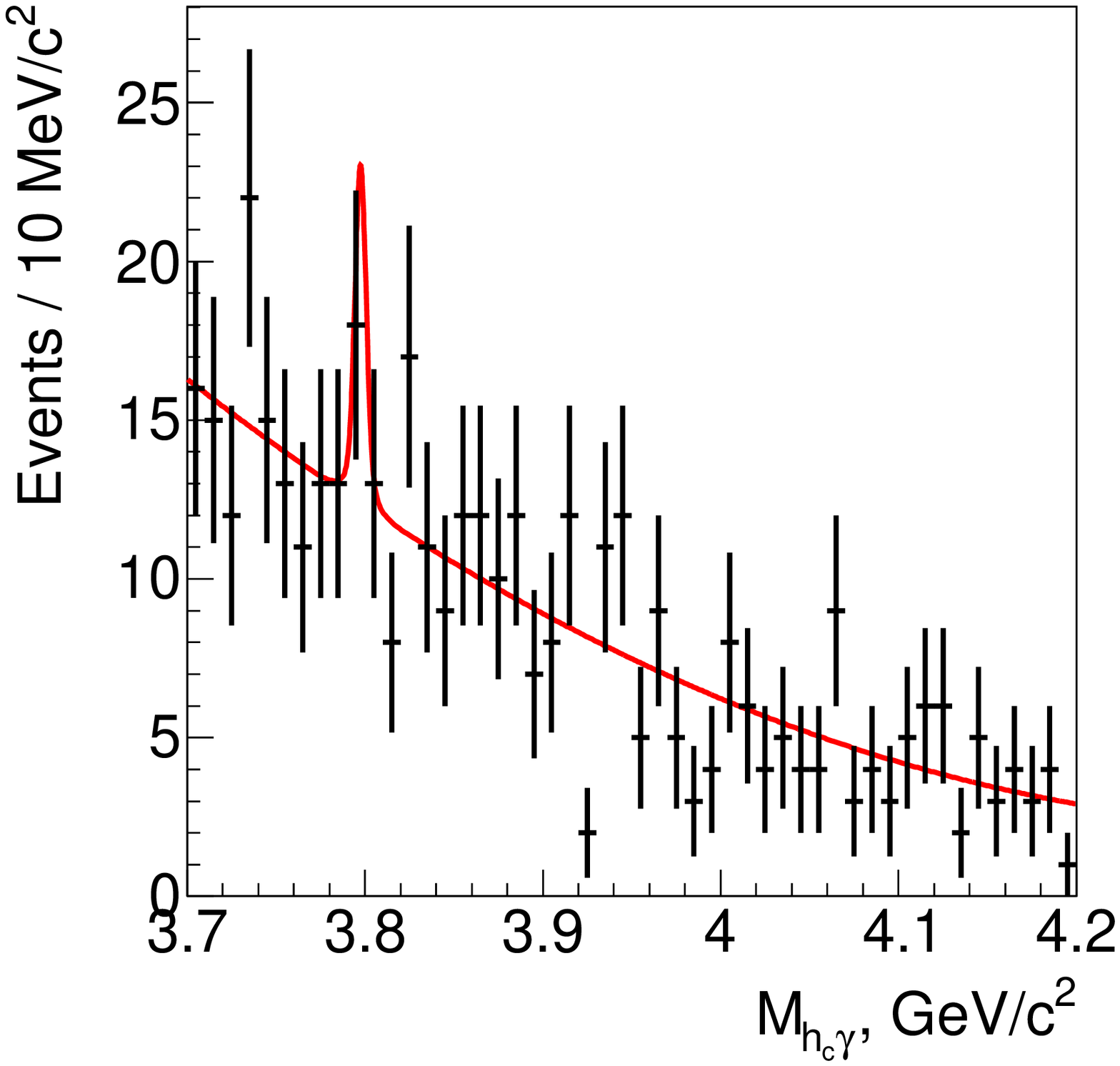}
\includegraphics[width=7cm]{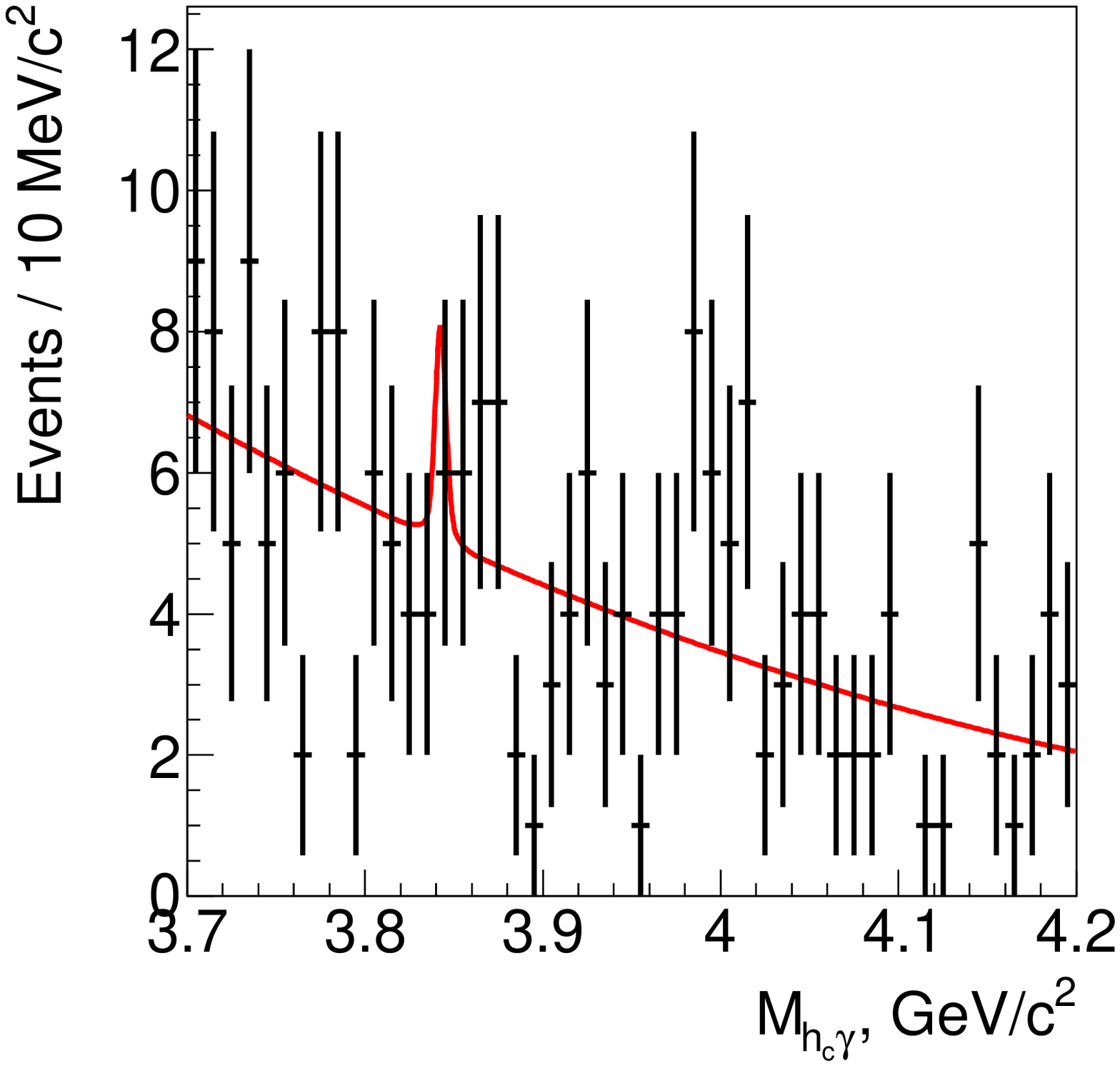} \\
\caption{Fit results for the channels $\decaypimkp$ (left) and
$\decaypimks$ (right).}
\label{fig:etac2pimk_fit_data}
\end{figure}

The fit results corresponding to the most significant peaks within the
$\eta_{c2}(1D)$ search region are shown in figure~\ref{fig:etac2kp_fit_data}
for the decays $\decaykp$ and $\decayks$,
in figure~\ref{fig:etac2k_fit_data} for their combination,
and in figure~\ref{fig:etac2pimk_fit_data} 
for the decays $\decaypimkp$ and $\decaypimks$.
The masses, yields, and local significances of the most significant peaks
within the search region are listed in table~\ref{tab:fit_results}.
The local significances are calculated from the difference of $(-2 \ln L)$
assuming that the mass is known (with one degree of freedom).
There is no significant signal in any channel;
since even the local significance does not exceed $3\sigma$, the global
significance is not calculated.

\begin{table}
\caption{The masses, yields, and local significances of the most significant
peaks within the search region.}
\begin{tabular}{c|c|c|c}
\hline\hline
Channel & Mass, $\mevcc$ & Yield & Local significance \\
\hline
$\decaykp$                & $3809.6 \pm 4.3$ & $3.3 \pm 3.0$ & $1.3\sigma$ \\
$\decayks$                & $3814.4 \pm 2.7$ & $2.7 \pm 2.3$ & $1.5\sigma$ \\
$\decaykp$ and $\decayks$ & $3821.8 \pm 4.4$ & $1.6 \pm 3.2$ & $0.7\sigma$ \\
$\decaypimkp$             & $3797.0 \pm 1.6$ & $9.4 \pm 5.1$ & $2.1\sigma$ \\
$\decaypimks$             & $3842.3 \pm 3.4$ & $2.6 \pm 3.1$ & $1.0\sigma$ \\
\hline\hline
\end{tabular}
\label{tab:fit_results}
\end{table}

\subsection{Systematic uncertainties}
\label{sec:systematic_uncertainties}

The systematic errors in the
branching-fraction products can be subdivided into three categories:
branching-fraction scale errors, resolution errors, and model errors.

The sources of the systematic uncertainty in the branching-fraction
scale include
overtraining (the difference between the efficiency in the training and testing
samples),
the error on the difference of the particle-identification requirement
efficiency between the data and MC,
the tracking efficiency error,
the difference between the MLP efficiency for data and MC,
the unknown amplitude of the $\decaypimkp$ and $\decaypimks$ decays,
the number of $\Upsilon(4S)$ events, and
the $\eta_c$, $h_c$, and $\Upsilon(4S)$ branching fractions.

The difference of the particle-identification requirement
efficiency between the data and MC is estimated from several control samples,
including $D^{*+}\to D^0(\to K^-\pi^+)\pi^+$ for $K$ and $\pi$,
$\Lambda \to p \pim$ for $p$,
$\Lambda_c^+ \to \Lambda \pip$ for $\Lambda$,
$\Dz \to \ks \pip \pim$ for $\ks$, and $\tau^- \to \pim \piz \nu_\tau$ for
$\piz$ candidates, respectively.

The uncertainty due to the difference in the MLP efficiency between the data and
MC is estimated using the decay mode $\Bz \to \eta_c \pim \kp$. This decay is
reconstructed using selection criteria that are as similar as possible to the
signal mode $\decaykp$. The MLP optimized for $\decaykp$ is applied to
the control channel. Some MLP input variables used for the signal channel are
undefined for $\Bz \to \eta_c \pim \kp$, for example, the $\piz$ likelihood
for photons. Such variables are set to constants. The ratio of the
number of signal candidates in all channels after the application of MLP
selection and in the channel $\etacdec{0}$ before the application of MLP
selection is extracted from a simultaneous fit to the
$\eta_c$ mass distributions and compared to its value in MC.
The control-channel events are weighted to reproduce the selection
efficiencies in the signal channel.
The resulting ratio of the number of $\eta_c$ candidates is
$0.82 \pm 0.10$, while the ratio in MC is $0.98$.
The relative difference between the data and MC efficiency is 16.8\%
and the statistical error of its determination is 9.9\%.
The statistical error is also added in quadrature to the systematic uncertainty.
The resulting uncertainty from the MLP selection
efficiency difference in data and MC is 19.5\%. Since only
the channel $\etacdec{0}$ is used without the MLP selection, this error
includes also the error of the branching fractions of other $\eta_c$ channels
relative to the channel $\etacdec{0}$.

The MLP efficiency uncertainty does not include
the uncertainty caused by the difference between the data and MC in
the distributions of the variables that are not defined for the channel
$\Bz \to \eta_c \pim \kp$. There are six such variables:
the $\eta_{c2}(1D)$ helicity angle, the $h_c$ daughter-photon azimuthal angle,
the $h_c$ and $\eta_c$ masses, and the $\piz$ likelihoods of the $\eta_{c2}(1D)$
and $h_c$ daughter photons. The distributions of the angular variables
are known assuming negligible contribution of higher multipole amplitudes.
No additional systematic uncertainty for the difference of
their distributions in data and MC is assigned.

The error due to the $\eta_c$ mass
distribution uncertainty is estimated by varying the $\eta_c$ mass and width
by $\pm 1 \sigma$ and reweighting the selected MC events; the largest
resulting efficiency difference is treated as the systematic uncertainty
from the $\eta_c$ mass distribution. The $\eta_c$ width uncertainty
is increased up to the difference of the $\eta_c$ input and measured widths
in the channel $\Bz \to \eta_c \pim \kp$ to take into account
the possible difference of the resolution.

Since the $h_c$ has a daughter photon that is not included into any
kinematic fits before the calculation of the $h_c$ mass, the uncertainty
in the $h_c$ mass distribution is caused mostly by the difference of
the resolution in the photon energy in data and MC. This uncertainty
is estimated by varying the photon energy correction~\cite{Tamponi:2015xzb} by
$\pm 1\sigma$, reconstructing the MC events again using the new correction,
and calculating the difference between the resulting efficiencies.

The uncertainty associated with the photon $\piz$ likelihoods is estimated using
the decay $B^+ \to \psi(2S) (\to \chi_{c1} (\to J/\psi \gamma) \gamma) K^+$.
A neural network consisting of only two likelihoods is used to select the
events in data and MC. The number of $\psi(2S)$ events in the data is calculated
from a fit to the $(\chi_{c1} \gamma)$ invariant mass
both before and after the application of the MLP selection. The
difference of the efficiencies in data and MC is found to be 4.6\%.

The uncertainty related to the unknown amplitude of the $\decaypimkp$ and
$\decaypimks$ decays is estimated by considering several decay
amplitudes. By default, the distribution is assumed to be uniform.
As an alternative, the decay is assumed to proceed via the intermediate
$K^*(892)$ resonance. Two possible $K^*(892)$ polarizations are
considered: $\lambda_{K^*(892)} = \pm 1$ and $\lambda_{K^*(892)} = 0$, where
$\lambda_{K^*(892)}$ is the $K^*(892)$ helicity. The angular distribution
of the $K^*(892)$ decay is given by
$|d^1_{\lambda_{K^*(892)}\,0}(\theta_{K^*(892)})|^2$, where
$d$ is the Wigner $d$-function, and $\theta_{K^*(892)}$ is the $K^*(892)$
helicity angle (the angle between $-\vec{p}_{\eta_{c2}(1D)}$ and
$\vec{p}_K$, where $\vec{p}_{\eta_{c2}(1D)}$ and $\vec{p}_K$ are the momenta of
the $\eta_{c2}(1D)$ and $K$ in the $K^*(892)$ rest frame, respectively).
The maximal deviations of the efficiency for alternative amplitude models
from the default one are considered as systematic uncertainty.
The uncertainties are found to be 15.0\% and 9.5\% for $\decaypimkp$
and $\decaypimks$, respectively.

All systematic errors related to the branching-fraction scale
are listed in table~\ref{tab:branching_scale_error}.
The errors of the tracking efficiency and the difference of
the particle-identification requirement efficiency depend on the $h_c$ decay
channel; the values presented in table~\ref{tab:branching_scale_error}
are weighted averages.

\begin{table}
\caption{The systematic uncertainties of the branching-fraction scale.}
\centering
\begin{tabular}{c|c|c|c|c}
\hline\hline
Error source & $\eta_{c2}(1D) \kp$ & $\eta_{c2}(1D) \ks$ & $\eta_{c2}(1D) \pim \kp$ & $\eta_{c2}(1D) \pim \ks$ \\
\hline
Overtraining & 3.7\% & 4.2\% & 2.0\% & 3.4\% \\
PID & 4.2\% & 5.1\% & 5.0\% & 5.5\% \\
Tracking & 1.5\% & 1.9\% & 1.9\% & 2.2\% \\
MLP efficiency & 19.5\% & 19.5\% & 19.5\% & 19.5\% \\
$\piz$ likelihoods & 4.6\% & 4.6\% & 4.6\% & 4.6\% \\
$\eta_c$ mass distribution & 5.5\% & 5.3\% & 5.6\% & 5.6\% \\
Photon energy resolution & 2.5\% & 1.7\% & 2.7\% & 1.3\% \\
Amplitude of $\decaypimk$ & --- & --- & 15.0\% & 9.5\% \\
Number of $\Upsilon(4S)$ events & 1.4\% & 1.4\% & 1.4\% & 1.4\% \\
$\br$ of $\eta_c$ and $h_c$ & 13.6\% & 13.6\% & 13.6\% & 13.6\% \\
$\br(\Upsilon(4S) \to B \bar{B})$ & 1.2\% & 1.2\% & 1.2\% & 1.2\% \\
\hline
Total & 25.7\% & 25.9\% & 29.8\% & 27.6\% \\
\hline\hline
\end{tabular}
\label{tab:branching_scale_error}
\end{table}

\subsection{Branching fraction}

Since no significant signal is observed, a mass scan is performed over the
$\eta_{c2}(1D)$ search region with a step size of $0.5\ \mevcc$.
The confidence intervals
for the branching-fraction products are calculated at each point.

The resolution scaling coefficient $\mathcal{S}$ is measured by modifying
the resolution function $R_{\eta_{c2}(1D)}$:
\begin{equation}
R_{\eta_{c2}(1D)}(\Delta M) \to \frac{1}{\mathcal{S}}
R_{\eta_{c2}(1D)}\left(\frac{\Delta M}{\mathcal{S}}\right),
\end{equation}
and similarly for other processes.
The difference of the resolution in the data and MC is
estimated using the decay
$B \to \psi(2S) (\to \chi_{c1} (\to J/\psi \gamma) \gamma) K$.
This decay has two radiative transitions similar to the signal processes.
The selection of the control channel
is performed using a similar neural network,
which has the same photon-related variables as in the signal process.
After the photon resolution correction, no significant difference
is observed between the resolution in the $\chi_{c1}$ mass in data and MC.
The $\chi_{c1}$ mass resolution scaling coefficient
is found to be $(0.99 \pm 0.18)$ from a fit to the $\chi_{c1}$ mass.
However, the resolution in the $\psi(2S)$ mass in data is found to be worse
than the resolution in MC. The resolution scaling coefficient determined from
a fit to the $\chi_{c1} \gamma$ invariant mass distribution is
$(1.31 \pm 0.12)$.

Four resolution scaling coefficients are selected for analysis:
the nominal resolution ($\mathcal{S} = 1.00$), the scaling
coefficient determined from
$B \to \psi(2S) (\to \chi_{c1} (\to J/\psi \gamma) \gamma) K$
($\mathcal{S} = 1.31$), and the same result varied by $\pm 1 \sigma$
($\mathcal{S} = 1.19$ and $\mathcal{S} = 1.43$).

For each of the selected scaling coefficients, several signal models
are considered. They include the default model, the model without the signal,
the model with a higher-order (2 for $\decayks$ and 3 for other decays)
background polynomial, a model with variations of
the $M_{h_c \gamma}$ fitting region, and
a model with alternative values of the $\eta_{c2}(1D)$ width
($\Gamma_{\eta_{c2}(1D)} = 0\ \mev$ and $1\ \mev$).

The confidence intervals are calculated for each model taking the
branching-fraction scale error into account. For the channel $\decaypimkp$,
the yield and its error are determined from the fit. To take the systematic
error into account, the Feldman-Cousins unified confidence
intervals~\cite{Feldman:1997qc} for the
branching-fraction distribution are used:
\begin{equation}
\begin{aligned}
\br & = N_{\eta_{c2}(1D)} k, \\
\sigma_{\br} & =
  \sqrt{\sigma_{N_{\eta_{c2}(1D)}}^2 k^2 +
  N_{\eta_{c2}(1D)}^2 (k \sigma_{\br}^{\text{(scale)}})^2 +
  \sigma_{N_{\eta_{c2}(1D)}}^2 (k \sigma_{\br}^{\text{(scale)}})^2}, \\
\end{aligned}
\end{equation}
where $\br$ and $\sigma_{\br}$ are the mean and variance of the
branching-fraction distribution,
respectively, $k$ is the ratio of the branching fraction and observed number of
events, $\sigma_{N_{\eta_{c2}(1D)}}$ is the uncertainty in the $\eta_{c2}(1D)$
yield, and $\sigma_{\br}^{\text{(scale)}}$ is the relative
branching-fraction scale error
determined in section~\ref{sec:systematic_uncertainties}
(the total error from table~\ref{tab:branching_scale_error}).
Since the $\eta_{c2}(1D)$ yield is determined from the fit, the model without
the signal is excluded from the list of alternative models for $\decaypimkp$.

For other decays: $\decaykp$, $\decayks$, and $\decaypimks$,
it is not possible to determine the yield
at all masses from the fit, because the number of events is too small.
There are gaps without any events, where the likelihood is a continuously
increasing function of the signal yield for all allowed values of the yield
(such values that the overall fitting function is positive). Thus, it is
necessary to switch to event counting. The counting region is chosen to be
within $\pm 1.5 (\sigma_1 + \sigma_2) / 2$ from the current value of mass,
where $\sigma_1$ and $\sigma_2$ are the parameters of the narrow asymmetric
Gaussian used in the resolution fit. The expected number
of background events is determined from the fit. The
profile-likelihood-based intervals described in ref.~\cite{Rolke:2004mj}
are used. The confidence-interval calculation takes into account the
branching-fraction scale error.

\begin{figure}
\centering
\includegraphics[width=7cm]{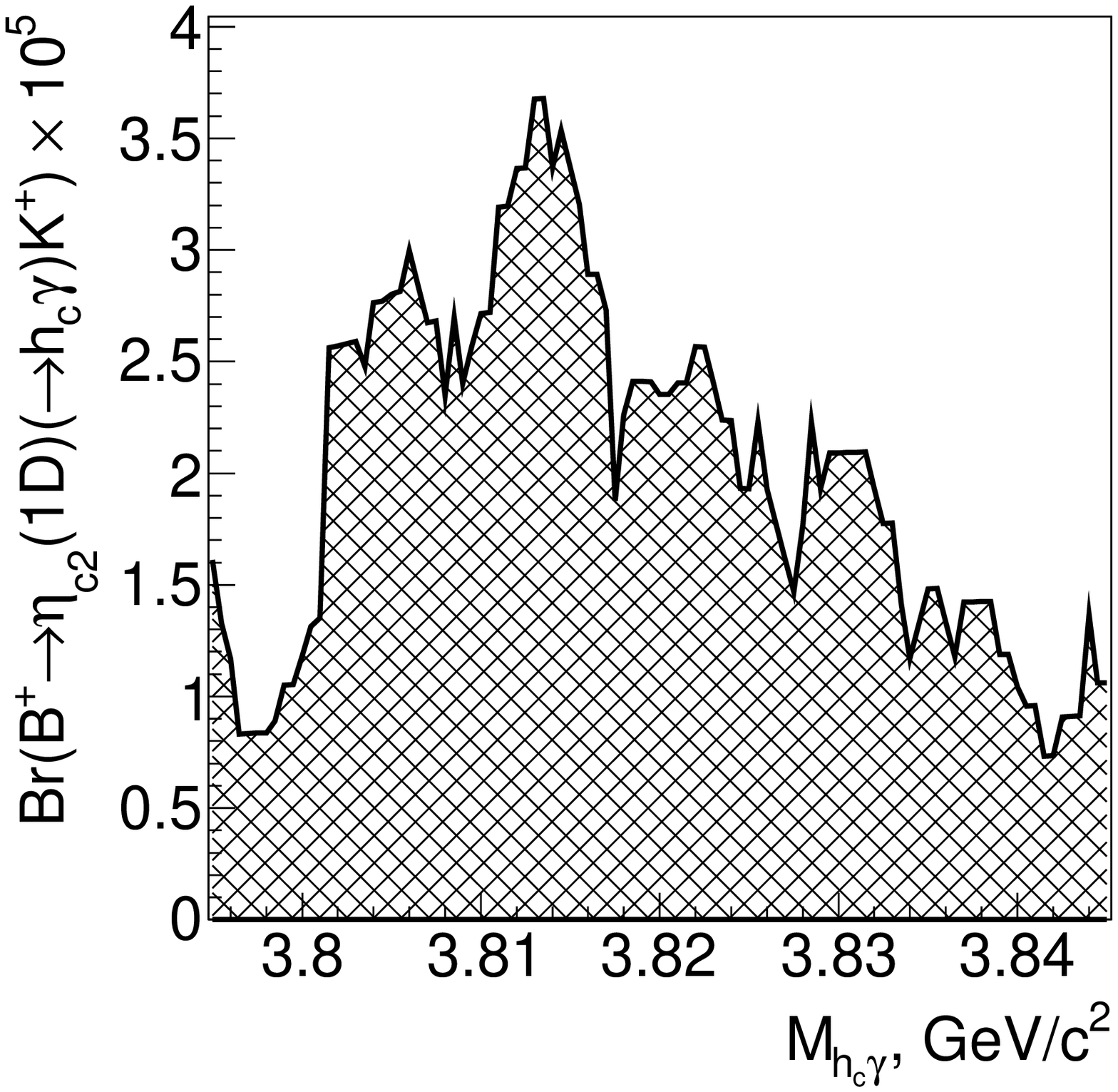}
\includegraphics[width=7cm]{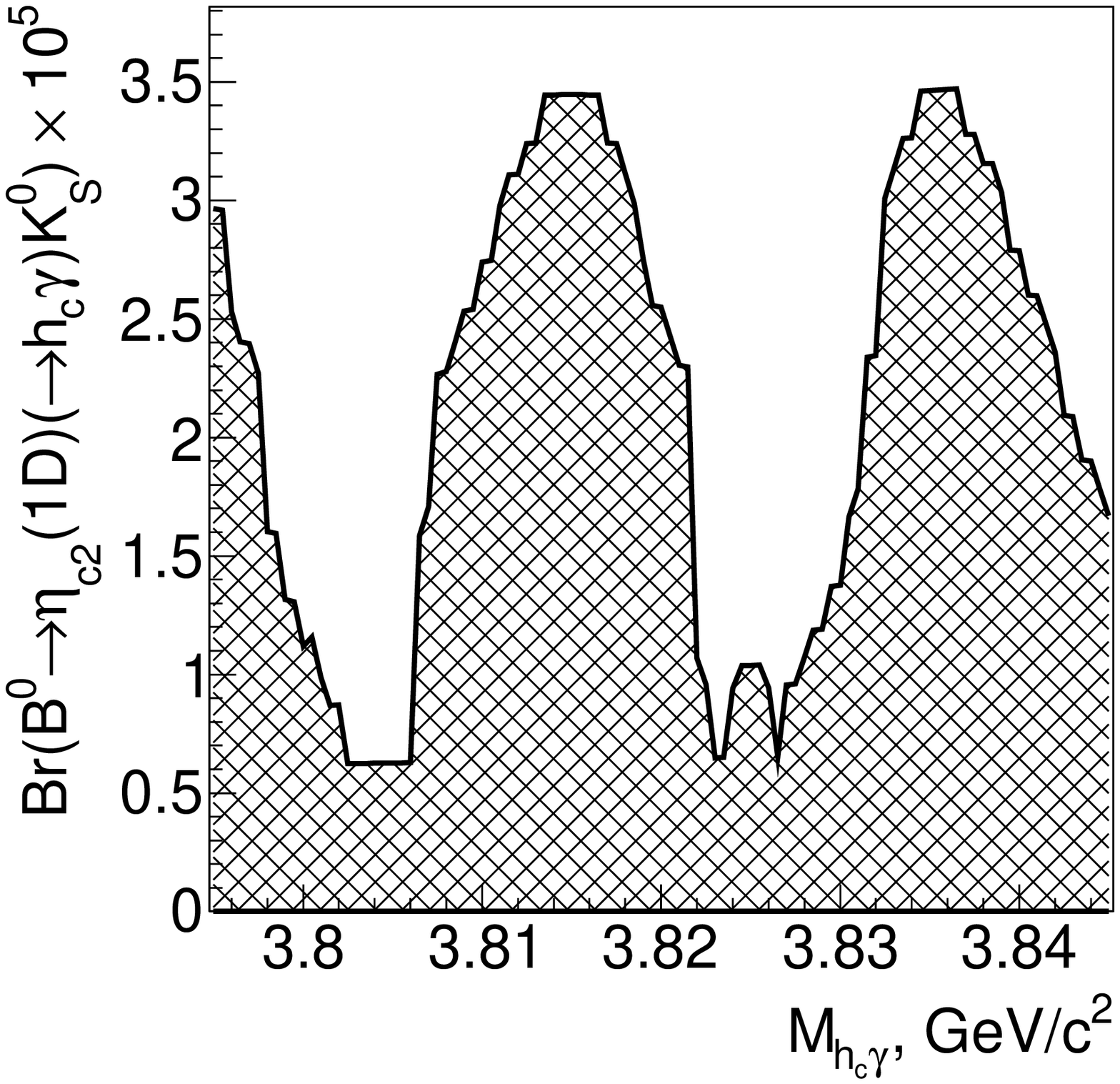}
\caption{Confidence intervals (90\% C. L.) for the branching-fraction products
for the channels $\decaykp$ (left) and $\decayks$ (right)
including the systematic uncertainties.}
\label{fig:etac2k_scan_systematics}
\end{figure}

\begin{figure}
\centering
\includegraphics[width=7cm]{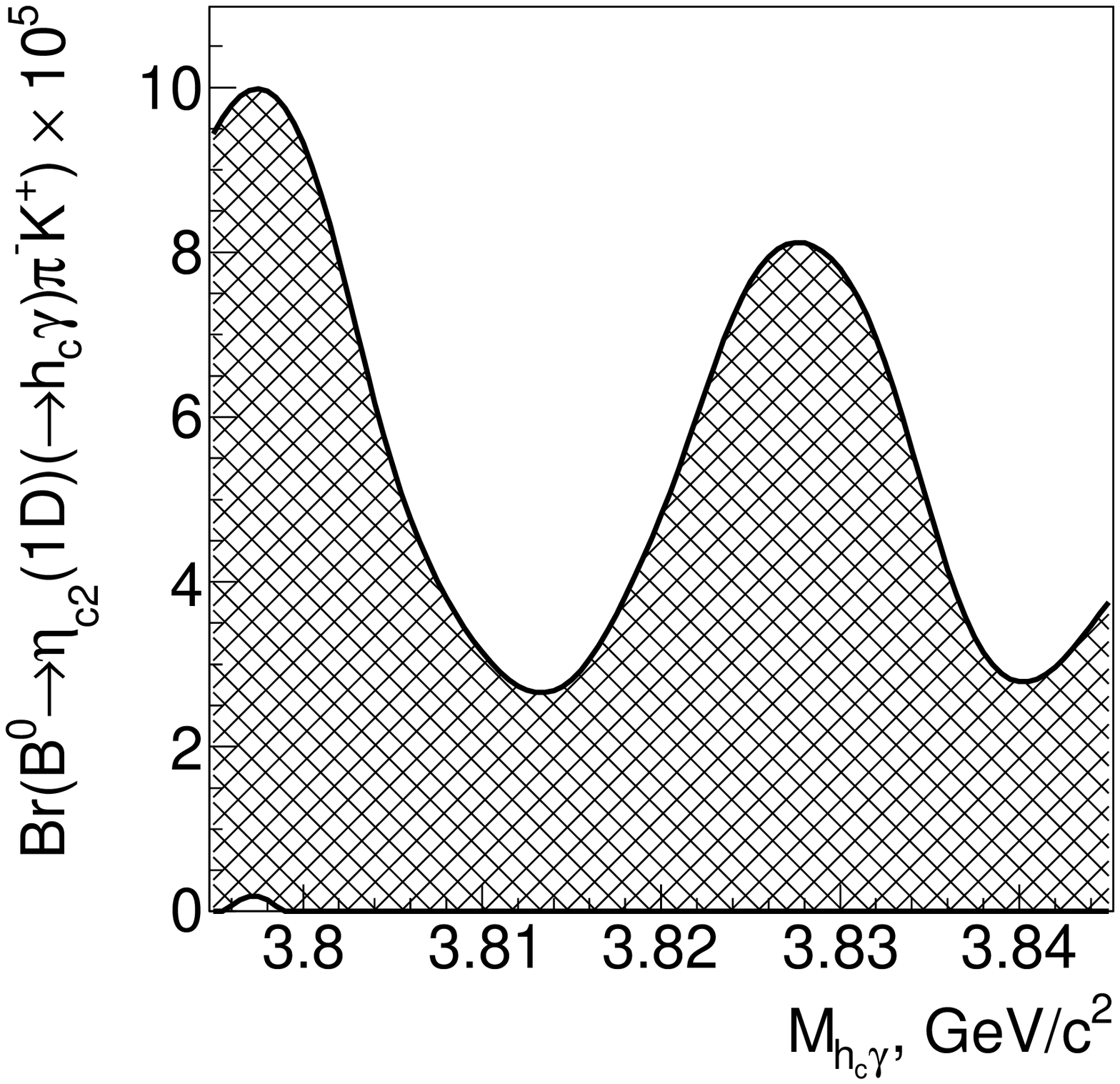}
\includegraphics[width=7cm]{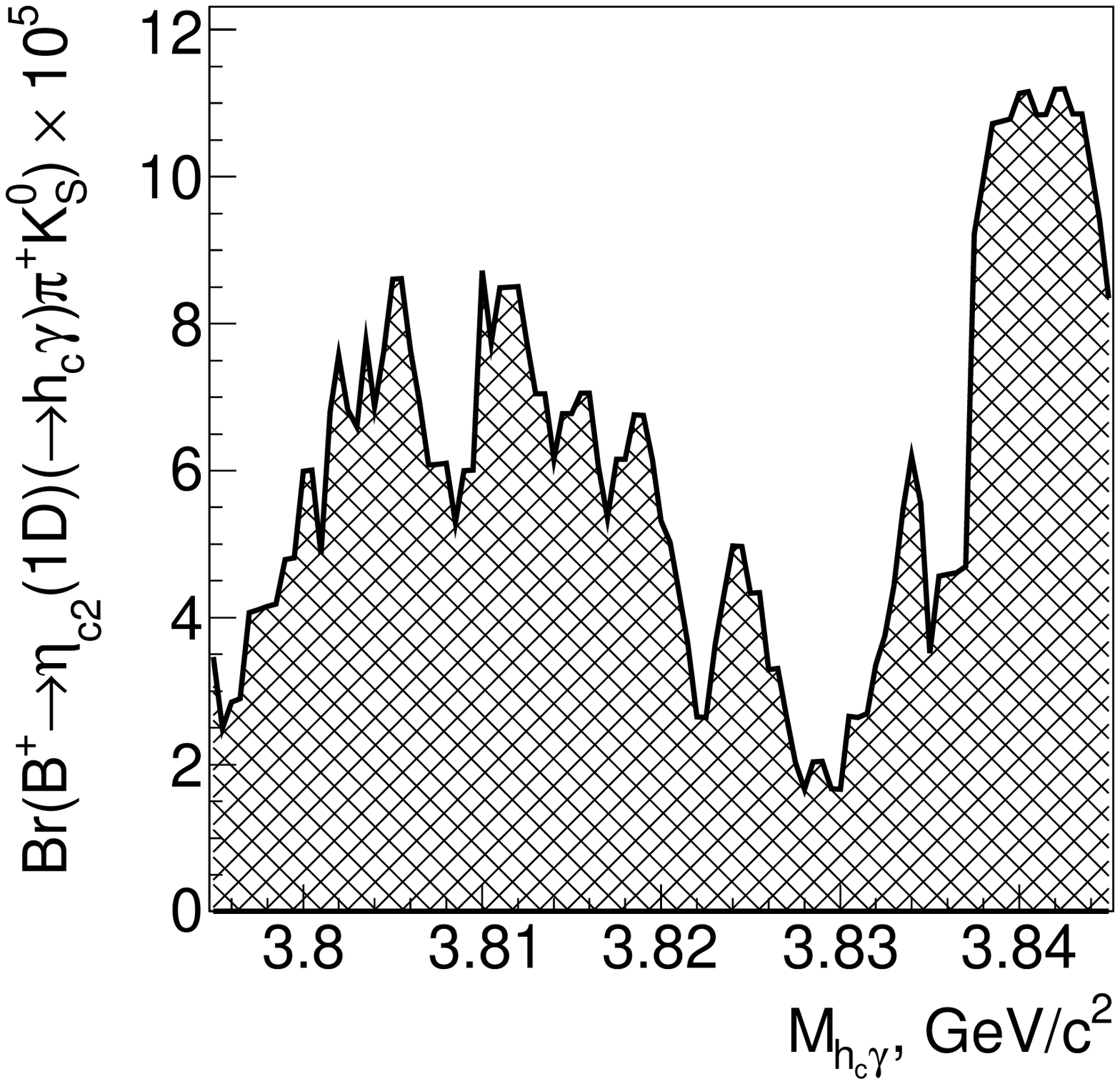}
\caption{Confidence intervals (90\% C. L.) for the branching-fraction products
for the channels $\decaypimkp$ (left) and $\decaypimks$ (right)
including the systematic uncertainties.}
\label{fig:etac2pik_scan_systematics}
\end{figure}

The results of the confidence-interval determination for all
resolution scaling coefficients and signal models are merged. For a
specific $M_{h_c \gamma}$, the minimal lower limit and the maximal upper limit
are selected. The resulting confidence intervals are shown in
figure~\ref{fig:etac2k_scan_systematics} for the channels $\decaykp$
and $\decayks$ and in figure~\ref{fig:etac2pik_scan_systematics}
for the channels $\decaypimkp$ and $\decaypimks$.

\section{Conclusions}

A search for the $\eta_{c2}(1D)$ using the decays
$\decaykp$, $\decayks$, $\decaypimkp$ and $\decaypimks$
has been carried out.
No significant signal is found. Confidence intervals
for branching fractions are determined in the $\eta_{c2}(1D)$
search range from 3795 to 3845 $\mevcc$. The scan results are shown in
figure~\ref{fig:etac2k_scan_systematics} and
figure~\ref{fig:etac2pik_scan_systematics}. The upper limits at 90\% C. L.
corresponding to $\eta_{c2}(1D)$ masses within the search range are
\begin{equation*}
\begin{aligned}
\br(\decaykp) \times \br(\eta_{c2}(1D) \to h_c \gamma) &
  < 3.7 \times 10^{-5}, \\
\br(\decayks) \times \br(\eta_{c2}(1D) \to h_c \gamma) &
  < 3.5 \times 10^{-5}, \\
\br(\decaypimkp) \times \br(\eta_{c2}(1D) \to h_c \gamma) & 
  < 1.0 \times 10^{-4}, \\
\br(\decaypimks) \times \br(\eta_{c2}(1D) \to h_c \gamma) &
  < 1.1 \times 10^{-4}. \\
\end{aligned}
\end{equation*}
The measured upper limit for
$\br(\decaykp) \times \br(\eta_{c2}(1D) \to h_c \gamma)$ is
consistent with the existing theoretical prediction of
$(1.72 \pm 0.42) \times 10^{-5}$~\cite{Xu:2016kbn}.
A data sample of about $10\ \text{ab}^{-1}$ is required to reach the
expected value of branching-fraction product
$\br(\decaykp) \times \br(\eta_{c2}(1D) \to h_c \gamma) \sim
1.0 \times 10^{-5}$. Thus, the Belle II experiment should be able to observe
the $\eta_{c2}(1D)$ or exclude the predicted branching fraction
in the future~\cite{Kou:2018nap}.

\section*{Acknowledgements}

We thank the KEKB group for the excellent operation of the
accelerator; the KEK cryogenics group for the efficient
operation of the solenoid; and the KEK computer group, and the Pacific Northwest National
Laboratory (PNNL) Environmental Molecular Sciences Laboratory (EMSL)
computing group for strong computing support; and the National
Institute of Informatics, and Science Information NETwork 5 (SINET5) for
valuable network support.  We acknowledge support from
the Ministry of Education, Culture, Sports, Science, and
Technology (MEXT) of Japan, the Japan Society for the 
Promotion of Science (JSPS), and the Tau-Lepton Physics 
Research Center of Nagoya University; 
the Australian Research Council including grants
DP180102629, 
DP170102389, 
DP170102204, 
DP150103061, 
FT130100303; 
Austrian Science Fund (FWF);
the National Natural Science Foundation of China under Contracts
No.~11435013,  
No.~11475187,  
No.~11521505,  
No.~11575017,  
No.~11675166,  
No.~11705209;  
Key Research Program of Frontier Sciences, Chinese Academy of Sciences (CAS), Grant No.~QYZDJ-SSW-SLH011; 
the  CAS Center for Excellence in Particle Physics (CCEPP); 
the Shanghai Pujiang Program under Grant No.~18PJ1401000;  
the Ministry of Education, Youth and Sports of the Czech
Republic under Contract No.~LTT17020;
the Carl Zeiss Foundation, the Deutsche Forschungsgemeinschaft, the
Excellence Cluster Universe, and the VolkswagenStiftung;
the Department of Science and Technology of India; 
the Istituto Nazionale di Fisica Nucleare of Italy; 
National Research Foundation (NRF) of Korea Grant
Nos.~2016R1\-D1A1B\-01010135, 2016R1\-D1A1B\-02012900, 2018R1\-A2B\-3003643,
2018R1\-A6A1A\-06024970, 2018R1\-D1A1B\-07047294, 2019K1\-A3A7A\-09033840,
2019R1\-I1A3A\-01058933;
Radiation Science Research Institute, Foreign Large-size Research Facility Application Supporting project, the Global Science Experimental Data Hub Center of the Korea Institute of Science and Technology Information and KREONET/GLORIAD;
the Polish Ministry of Science and Higher Education and 
the National Science Center;
the Russian Foundation for Basic Research Grant No.~18-32-00277;
University of Tabuk research grants
S-1440-0321, S-0256-1438, and S-0280-1439 (Saudi Arabia);
the Slovenian Research Agency;
Ikerbasque, Basque Foundation for Science, Spain;
the Swiss National Science Foundation; 
the Ministry of Education and the Ministry of Science and Technology of Taiwan;
and the United States Department of Energy and the National Science Foundation.

\end{document}